# Age-stratified clustering of multiple long-term conditions

Anirban Chakraborty[1,2], Bruce Guthrie[3], Sohan Seth[1 3]

1. School of Informatics, University of Edinburgh UK
2. School of Engineering, Computing and Mathematical Sciences, University of Wolverhampton
3. Advanced Care Research Centre, Usher Institute, University of Edinburgh UK

Corresponding author:

Sohan Seth, School of Informatics, University of Edinburgh, 10 Crichton Street, Edinburgh EH8 9AB,

Sohan.Seth@ed.ac.uk

Abstract

**Background:** Most people with any long-term condition have multiple long-term conditions, but our understanding of how conditions cluster is limited. Many clustering studies identify clusters in the whole population, but the clusters that occur in people of different ages may be distinct. The aim of this paper was to explore similarities and differences in clusters found in different age-groups.

**Method:** We present a method for finding similar clusters in multiple age-groups, referred to as cluster sets, using Latent Class Analysis (LCA) and Chebyshev distance metric. We analyse a primary care electronic health record (EHR) dataset recording the presence of 40 long-term conditions (LTCs) in 570,355 people aged 40-99 years with at least one of these conditions, analysing in five-year age-groups.

**Findings:** We find that the 600 clusters found separately in 12 age-strata can be summarised by 342 cluster sets with 263 cluster sets only being found in a single age-group (singleton cluster sets), and 79 cluster sets being present in multiple age-groups. We observe that 31 conditions of the 40 conditions studied appear in cluster sets with the respective condition being the only condition present with a very high prevalence of more than 0.9 whereas the remaining cluster sets typically contain two to four conditions present with a high prevalence of more than 0.7.

**Interpretation:** Multimorbidity profiles in different age-groups are often distinct (singleton cluster sets observed only in that age-group), but similar clusters with small variations in their composition are also found in multiple age-groups. This demonstrates the age dependency of MLTC clusters and presents a case for age-stratified clustering.

**Funding:** National Institute for Health and Care Research AIM-CISC programme (NIHR202639).

## Introduction

Multimorbidity, the co-occurrence of two or more long-term conditions (LTCs) in an individual, is becoming increasingly prevalent due to rapid population ageing worldwide. Multimorbidity is associated with poorer health outcomes, increased healthcare utilization, and higher healthcare costs,[1,2] and therefore, identifying individuals with similar multimorbidity profiles has the potential to inform effective prevention and management[3]. Clustering analysis is an increasingly used approach for identifying multimorbidity profiles where individuals are grouped in distinct clusters, and clusters are represented by their composition (for example, a cluster can have a high probability of anxiety and depression and a low probability of hypertension).

The experience of multimorbidity varies by age, with younger people typically having fewer and more obviously closely related disorders, such as mental health conditions and substance abuse, while older people more often experience more complex combinations of physical conditions and mental health conditions of old age such as dementia[4]. Many clustering studies examine clusters in whole populations without stratifying by age, while a smaller number of studies explore clusters in broad age-strata such as young (18-44 years), middle-aged (45-64), older (65-84) and very old (85+) adults[5,6,7,8,9,10,11,12,13,14,15,16,17].

There are good reasons why condition clustering may vary by age-group, meaning that whole population clustering may be misleading. This is because observations of LTCs are censored at the current age and require survival to that age. We therefore do not have knowledge of conditions that an individual might develop in the future. For example, consider two people who only have hypertension aged 20 and aged 80 years. The younger person is an outlier (hypertension is rare in young adults) and has much longer to develop complications of hypertension or other conditions than the older person, whereas hypertension is very common in 80-year-olds (and to survive to age 80 also requires not being diagnosed with earlier life-shortening conditions). However, when treated as members of a single population without age-stratification, they are similar because they are both living with only hypertension. The aim of this study was to examine the extent to which identified condition clusters varied between age-groups versus were observed in multiple, consecutive age-groups.

## Methods

### Overall study design

The overall design is to use latent class analysis to examine condition clusters in the whole population of adults aged 40-99 years and in five-year age strata, and to then identify clusters which are present in consecutive age-groups. We expected a priori that some clusters would only be found in one age-group ('singleton cluster sets') whereas other clusters inferred in different age-groups would be related ('non-singleton cluster sets'). Clusters in a non-singleton cluster set are not necessarily identical since there may be a gradual shift in composition over age groups. The primary methodological contribution of this paper is to develop a protocol for identifying MLTC cluster sets from individuals stratified by age (Figure 1).

### Dataset

Our study selected 570,355 adults aged 40-99 with at least one long-term condition from an existing multimorbidity dataset[1] with data for approximately 1.75 million patients of all ages, representing approximately one-third of the Scottish population (Table 1). The data was collected on March 21, 2007 from 314 medical practices and is representative of the broader Scottish population. The dataset has several strengths, including its high level of representativeness and the use of comprehensive EHR data to define forty purposively selected LTCs. The NHS National Research Ethics Service had previously approved the anonymous use of these data for research purposes, therefore this study did not need individual ethics approval.

### Age Stratification for clustering

Multimorbidity is very strongly associated with age. As individuals age, the number of LTCs they accrue and the proportion of people with MLTCs rapidly rises. By age 40, approximately one in three individuals have at least one long-term condition and about 15% have multimorbidity, and in over-65 years old multimorbidity is the norm[1]. Given this understanding, our investigation focuses on individuals aged 40 to 99 years, which includes the period of adulthood when multimorbidity is most common, and the period when most people transition from single conditions to multimorbidity. While there are various options for age stratification in clustering[4,5], we chose to use a 5-year stratification to strike a balance between over-granularity where each group might have too few individuals to generate reliable clusters, and under-granularity where subsequent age-groups might have very different clusters that cannot be aligned reliably. Specifically, we split the data into G = 12 subgroups with patients of age in years between G1:40-44, G2:45-49, ..., G12:95-99.

### Finding Clusters

LCA[18] is considered a useful approach for clustering long-term conditions because it has good stability and reproducibility compared to other commonly used methods such as k-medoids and hierarchical clustering[19]. LCA is a finite mixture model that, with the number of clusters or mixtures specified, assumes that a sample $y_i$ belongs to one of these clusters exclusively, and estimates the parameters $\theta_k$s of these clusters by maximizing the likelihood (the underlying parameters can also be estimated using Bayesian inference[20]),

$$\prod_{i=1}^{N} f(y_i | \Theta) = \prod_{i=1}^{N} \sum_{k=1}^{K} \pi_k \, p(y_i | \theta_k).$$

Here $\mathbf{y}$ is a $D$-dimensional binary vector denoting the presence and absence of $D$ conditions, $K$ is the number of clusters, $\boldsymbol{\theta}_k = [\theta_{1k}, \ldots, \theta_{dk}, \ldots, \theta_{Dk}]^T$ is a vector of probability values representing the probability of the $d$-th condition being present in the $k$-th cluster, $\pi_k$ is the prevalence of the $k$-th cluster, and $p(.)$ is the density function of a multivariate Bernoulli distribution. For each $K$, we have run the optimization 50 times to select the best solution in terms of BIC value[21]. We found the LCA solutions to be stable over different runs and different $K$ values. We set the number of clusters for each age-groups to 50 to strike a balance between comprehensiveness and complexity. We expect that if $K$ is too small then the inferred clusters will not be specific, i.e., each cluster will have a large number of conditions that are prevalent, while if $K$ is too large then each cluster will have fewer samples, thus, making the estimated parameters unreliable. Thus, each LCA run generates a $\theta$ matrix of dimension $40 \times 50$ for each of the 12 age-groups (e.g., see Supplement Figures 1-3).

## Finding Similar Clusters

To study if the clusters in the subsequent age-groups are related, we first align the clusters found in one age-group with similar clusters found in its adjacent age-group. We compare two parameter matrices, say $\theta_1$ and $\theta_2$ (both of dimension $D \times K$) corresponding to two subsequent age-groups, respectively. Note that each column $\boldsymbol{\theta}_k$ of the parameter matrix represents a cluster. We compute the similarity between two clusters $\theta_i^{(1)}$ and $\theta_j^{(2)}$ as

$$s(\boldsymbol{\theta}_i, \boldsymbol{\theta}_j) = 1 - d_c(\boldsymbol{\theta}_i, \boldsymbol{\theta}_j)$$

(2)

where $d_c$ is the Chebyshev distance defined as

$$d_c(\boldsymbol{\theta}_i, \boldsymbol{\theta}_j) = \max_{l \in \{1, \ldots, L\}} |\theta_{li} - \theta_{lj}|$$

(3)

We explored several similarity/distance measures[22], including cosine similarity, $\ell_2$ distance, Manhattan distance, and the key motivation for favouring Chebyshev distance lies in its sensitivity to outliers. Chebyshev distance computes the maximum difference between two parameter vectors over all dimensions, i.e., conditions in our case, rather than the average difference as done by the other distance metrics. Therefore, two clusters are similar in terms of Chebyshev distance when they are similar over all conditions compared to when they can be very different in some of the conditions but this information in averaged out in the context of other metrics. This characteristic is particularly valuable when striving to align the best cluster among all the clusters available in the next age-group.

## Aligning Similar Clusters

Our approach is to align clusters in each group with clusters in its previous group as best as possible. To achieve this, we employ a greedy approach using the $K \times K$ similarity matrices for each subsequent age-group pairs with entries $s(\boldsymbol{\theta}_i, \boldsymbol{\theta}_j)$, by aligning the most similar cluster pairs in both age-groups first, and then move to align the next most similar and so on. In particular, we begin by aligning cluster $\boldsymbol{\theta}_i^{(g)}$ of the $g$-th age-group, to cluster

$\boldsymbol{\theta}_i^{(g-1)}$ of its preceding group, i.e., $(g-1)$-th, where $s(\boldsymbol{\theta}_i, \boldsymbol{\theta}_j)$ is the highest similarity score among all possible cluster pairs, i.e, $K \times K$ of them. We then proceed to align the cluster pairs with the second highest similarity along the rest of the $(K-1) \times (K-1)$ pairs, and so on until a minimum threshold of similarity is reached at which point we stop the process. We set this threshold to 0.7, i.e., to align two clusters we allow the probability of any condition in the two clusters to differ by maximum 0.3. While comparing two groups, we again follow a greedy approach by starting from the lowest age-group. We start by comparing the first two age-groups (i.e., $G1$ and $G2$). Once all the clusters of the second group are aligned (and re-indexed) to the clusters of the first group, we align (and re-index) all the clusters of $G3$ to the aligned clusters of $G2$, and so on. Supplement Figure 4 shows the outcome of this alignment and re-indexing in form of a network with clusters as nodes, and similarity as edges when the similarity is more than 0.7.

Results

The study population was the 570,355 people aged 40 years and over in the dataset who had at least one of the 40 LTCs. Table 1 shows the distribution of participants by age, and the prevalence of the forty conditions in the study population. Seven conditions had more than 10% prevalence (Hypertension, Painful Condition, Depression, CHD, Dyspepsia, Diabetes, and Thyroid Disorders), twelve had prevalence between 5% and 10%, and the remaining 21 had prevalence <5%.

We aligned 600 clusters over all age-groups to find 342 cluster sets, of which 263 were singleton (only observed in one age-group) and 79 are non-singleton (observed in consecutive age-groups). The 79 non-singleton cluster sets were observed in between two and twelve age-groups, and most were dominated by a single condition. The 263 singleton cluster sets (i.e. those which were not closely related to a similar cluster in another age-group) showed considerable diversity in their composition and were more commonly identified in older age-groups. Figures 2 and 3 show all non-singleton cluster sets identified, and Supplementary Figures 5 and 6 show all singleton clusters.

We named cluster sets based on the composition of the first cluster in the set (i.e. the index cluster set observed in the youngest age-group that the cluster set appears in). We considered a condition to be of moderate prevalence within a cluster if its prevalence in the cluster set was between 30% and 70%. We considered the condition to be of high prevalence if its prevalence was >70%, and of low prevalence if its prevalence was <30%. There were 47 (12 singleton and 35 non-singleton) cluster sets which had only a single condition that was highly prevalent with rest of the conditions being present at low prevalence (Figure 4 and Supplement Table 1). There were 36 (26 singleton and 10 non-singleton) cluster sets with one moderately prevalent condition (Supplement Table 2), 36 (13 singleton and 23 non-singleton) with two moderately prevalent conditions (Supplement Table 3), and 270 (49 singleton and 221 non-singleton) with three or more moderately prevalent conditions (Supplement Tables 4, 5, 6, 7).

Thirty one conditions out of forty had cluster sets with only the respective condition with high prevalence and no moderately prevalent conditions (i.e., every other condition only appeared with <0.3 prevalence) (Figure 4). These cluster sets can have different lengths with (EatingDisorders) (S54), (StrokeTIA) (S91) and (AtrialFibrillation) (S298) cluster sets (not an exclusive list) being singleton cluster sets present in G1 (40-44 years) only. The S40 (Hypertension) cluster set was the only one that was present in all age-groups (G1-G12, 40-99 years) and was the most prevalent non-singleton cluster set over the age-group. Other cluster sets present in all but the oldest age-groups were S52 (COPD) in G1-G11 (40-94 years) and S10 (Diabetes), S104 (DiverticularDisease), S211 (HearingImpairment), and S245 (Depression) all present in G1-G10 (40-89 years), and S312 (Dyspepsia) in G1-G9 (40-84 years). The S126 (Cancer) cluster set was present in G1-G7 (40-74 years) and was very similar to the S129 singleton cluster observed in G11 (80-84 years). Similarly, the S41 (InflammatoryArthritis) cluster set present in G1-G8 (40-79 years) was very similar to the S68 singleton cluster observed in G12 (95-99 years). Nine conditions (CKD, HeartFailure, Bronchiectasis, SchizophreniaBipolar, Dementia, VisualImpairment, ViralHepatitis, ChronicLiverDisease, and ParkinsonsDisease) did not have any cluster sets where they were the dominant conditions, but seven of them appeared in different cluster sets with high prevalence (CKD, ViralHepatitis, HeartFailure, SchizophreniaBipolar, Dementia, and VisualImpairment) or moderate prevalence (ChronicLiverDisease) alongside other conditions.

The prevalence in each age-group of individual conditions in the same non-singleton cluster set varied from very common (e.g., S40 (Hypertension) present in 6-13% of each age-group; S245 (Depression) in 2-11%) to rare (e.g., S280 (InflammatoryBowelDisease) in ~1%). The prevalence of conditions within non-singleton cluster sets varied by age-group. For example, prevalence within each age-group is weakly bimodal for S40 (Hypertension), continually declining for S245 (Depression), and increasing for S104 (DiverticularDisease). By definition, the single-condition dominated clusters had no other condition which is present in >30% of each age-group, but nonetheless there were small increases in prevalence of other conditions for clusters in older age-groups in most single-condition dominated cluster sets (Figure 4). In older age-groups, many but not all of the non-singleton cluster sets showed an increasing prevalence of Hypertension, e.g., S10 (Diabetes), S104 (DiverticularDisease), S245 (Depression), S312 (Dyspepsia), S41 (InflammatoryArthritis, S65 (PainfulCondition), and S74 (ThyroidDisorders) cluster sets.

Among the cluster sets with no moderately prevalent conditions in the first cluster, there were three non-singleton cluster sets with multiple highly prevalent conditions that are present in multiple age-groups. The S96 cluster set (Hypertension, Depression, Anxiety in the index cluster) that was present between age-groups G6-G10 (65-89 years) (Supplement Table 1) had increasing prevalence of CKD and StrokeTIA with increasing age, and the prevalence of the core conditions also increased with age. The S100 cluster set (PainfulCondition, InflammatoryArthritis) that was present between ages G3-G7 (50-74 years) has an increasing prevalence of CHD and Hypertension with age, but the prevalence of the core conditions decreased with age. There were cluster sets with a single dominant condition that appeared in younger age-groups but which re-appeared with moderately prevalent hypertension in older age-groups, e.g., S327 (MultipleSclerosis) appears in G1 (40-44 years) and S2 (MultipleSclerosis, Hypertension) appears in G8 (75-59 years).

There were a large number of cluster sets (singleton and non-singleton present in small numbers of age-groups) with complicated patterns of condition, although not uncommonly including as a core condition moderate prevalence Hypertension, Depression and PainfulCondition (which are the most three prevalent conditions in the whole population; Table 1). For example, S0 (SchizophreniaBipolar, Depression) is present in age-groups G1-G6 (40-69 years) with increasing prevalence of hypertension over age-groups. More complex combinations were usually only observed in single age-groups, although there were exceptions, including S131 (Depression, PainfulCondition, Asthma, IrritableBowelSyndrome).

To evaluate the difference between age-stratified clustering and whole-population clustering, we also identified 50 clusters in the whole population aged 40-99 years (Supplement Figure 7). We found some shared cluster sets in both approaches, such as single condition dominated HearingImpairment (W34), DiverticularDisease (W46), Dyspepsia (W19), Diabetes (W39), AlcoholProblem (W41), ThyroidDisorder (W50), IrritableBowelSyndrome (W31), and Cancer (W18). However, some single-condition dominated cluster sets identified in age-stratified clustering, such as S40 (Hypertension), S52 (COPD), or S41 (InflammatoryArthritis) were notably absent in the whole-population cluster sets despite being present as non-singleton clusters in all or almost all age-groups.

Discussion

We have developed a heuristic approach to explore the relatedness of MLTC clusters in different age-groups. Upon applying it on a dataset with 40 LTCs and 570,355 individuals aged over 40 and with at least one LTC, we observe various cluster sets that are present in multiple age-groups with slight variations (i.e., the prevalence of some conditions in non-singleton cluster sets can increase or decrease over the age-groups in the cluster set), or which are age-group specific. Notably, in any one age-group, the most frequent cluster sets were dominated by a single condition, and such cluster sets were commonly observed in many age-groups. However, when clustering in the whole population aged 40-99 years, single condition dominated cluster sets were less commonly observed. We expect the presence of non-singleton cluster sets, either because individuals follow similar common trajectories of disease accrual with different starting ages (hypertension will commonly be the first long-term condition diagnosed), or because some individuals who develop one or more diseases do not then rapidly accrue other conditions. The proposed approach provides a pseudo-temporal view of the progression of LTC clusters over age-groups in the population. A second notable feature of the data is that with increasing age, singleton clusters comprising multiple conditions become more frequent.

The proposed framework demonstrates that clusters found in different age-groups can be related. Although there are a considerable number of clusters that are age-group specific, there are also clusters that are present in different age-groups with only small variation in terms of their composition. This observation has several implications. First, clusters that are age-group specific indicate MLTC clusters that only affect a specific age-group, and show the diversity of multimorbidity profiles even in relatively close age-groups. Second, two seemingly unrelated clusters in widely separated age groups can be related when seen in the context of clusters observed in the age-groups in between. Instead of aligning the clusters in subsequent age-group, we can, first, cluster the population without age stratification, and then within each cluster, treat different age-groups separately, but we believe that this can be less effective as similar clusters in different age-groups can vary in composition. While our research design is inherently cross-sectional, the pursuit of identifying similar clusters across various age-groups lends a unique perspective akin to a longitudinal perspective (over age) within a cross-sectional experiment, and in some sense, identify progression of clusters in population than in individual. This is referred to as dynamic clustering where clusters are found in multiple temporally ordered datasets simultaneously under the assumption that their characteristics can change over time[23].

Although age-stratified clustering has been explored sporadically in the literature, the alignment of these clusters has not been investigated in detail, and existing studies usually age-stratify the population in broad age groups of 10 years or more and explore the characteristics of the resulting multimorbidity clusters in isolation. For example, age-stratified clustering with age groups <50, 50–64, 65–79, and ≥80 was explored alongside the association of these clusters to all-cause mortality[16]. Although some of the studies mention the similarity among clusters in different age-strata qualitatively, they do not do so quantitatively. For example, age- and sex-stratified clustering with age groups 65–79 and ≥80 years was explored and similar clusters were reported in different strata[10].

This is, to the best of our knowledge, the first study that explicitly explores similarities between age-stratified multimorbidity profiles. The proposed heuristic has several advantages methodologically including the use of LCA for clustering (which has been suggested to provide more stable clusters compared to other clustering approaches[19]), and the use of Chebyshev distance for estimating distance between clusters in subsequent age-groups, as this metric is more sensitive to change of prevalence of any condition in the clusters. Additionally, the dataset used is representative of the Scottish population and includes 40 common and important LTCs encompassing both physical and mental health conditions, and containing both LTCs that are highly prevalent and that are relatively rare.

The study also has some limitations. Although we have used LCA for clustering, any clustering method makes assumptions on how two objects are similar, and there is no definitive way to choose which approach is better. We assumed that there were the same number of clusters in each age-group which may also be restrictive as we might expect there to be more heterogeneity of LTCs in higher age-groups, although there are also fewer individuals which may restrict how reliably clusters can be identified. In this analysis, we chose to identify 50 cluster sets in both the whole population and in every age-group. Although some of the implied heterogeneity is reduced because there are many non-singleton clusters present in multiple age-groups, interpretation of so many clusters is necessarily difficult. However, analyses that choose to only identify very few clusters typically identify very heterogenous groups which are often 'everything' or catch-all clusters. The choice of threshold for forming a cluster set might also be improved further to be adaptive, as a single threshold might not be appropriate for all cluster sets. Cross-sectional analysis cannot distinguish whether similar clusters in different age-groups are because individuals develop similar patterns of condition but with onset at different ages (and then move out of that cluster as they develop other conditions) or because individuals tend to stay in a particular cluster without developing other conditions. Both are likely to be true, although we believe the former is likely to be more important.

The primary implication of this study is that whole population clustering is likely to be misleading, as from an individual perspective our data is censored at current age, while from a population perspective, multimorbidity clusters can vary over age groups with some clusters being unique to an age group while others are similar but not identical. In this context, longitudinal trajectories may be more realistic[24], although current studies usually do not explore the incompleteness of data. Our interpretation is that the single condition dominant clusters observed in multiple age-groups likely represent common entry points to MLTC (since there must be a first LTC experienced), with the more complex singleton clusters which dominate in older ages representing the end-point of different subsequent trajectories of accrual. One way of improving interpretation in longitudinal clustering may therefore be to stratify by both the first condition experienced (the start of the trajectory) and the age at which the first condition is experienced[25]. This requires developing more dedicated computational tools that can take into consideration the incompleteness of the multimorbidity trajectory alongside the different facets the data can be clustered on, e.g., are of onset and the first condition accrued.

## Conclusion

The standard approach of whole population clustering does not consider or reveal the nuances of MLTC clusters over age-groups, while the proposed approach of age-stratification and cluster set alignment explicitly shows the uniqueness and diversity of clusters in different age-groups as well as the presence of similar clusters with small variations in subsequent age-groups.

**Acknowledgements:** This work was funded by the National Institute for Health and Care Research (NIHR) Artificial Intelligence and Multimorbidity: Clustering in Individuals, Space and Clinical Context (AIM-CISC) grant NIHR202639. The views expressed are those of the author(s) and not necessarily those of the NIHR or the Department of Health and Social Care.

## Research in Context

**Evidence before this study.** Although age-stratified clustering has been explored sporadically in the literature, the alignment of these clusters has not been investigated in detail. We searched the PubMed database with keywords "multimorbidity", "clustering", ""age group"", and "stratified", for papers published between 2005 and 2025, and found 13 articles. These studies usually age-stratify the population in broad age groups of 10 years or more and explore the characteristics of the resulting multimorbidity clusters in isolation. Although some of the studies mention the similarity among clusters in different age-strata qualitatively, they do not do so quantitatively.

**Added value of this study.** To the best of our knowledge, this is the first study that explicitly looks at the alignment of clusters over different age groups. We stratify the population into granular groups of 5 years and propose a method to explicitly align multimorbidity clusters over consecutive age groups to explore the finer variation or the lack thereof of the respective multimorbidity clusters.

**Implications of all the available evidence.** We observe the age-dependency of multimorbidity clusters where similar clusters may exist in different age groups, but these clusters can also vary significantly, i.e., beyond a pre-defined threshold, over consecutive age groups. This implies that whole population clustering is likely to be misleading, while clusters of multimorbidity trajectories may be more realistic.

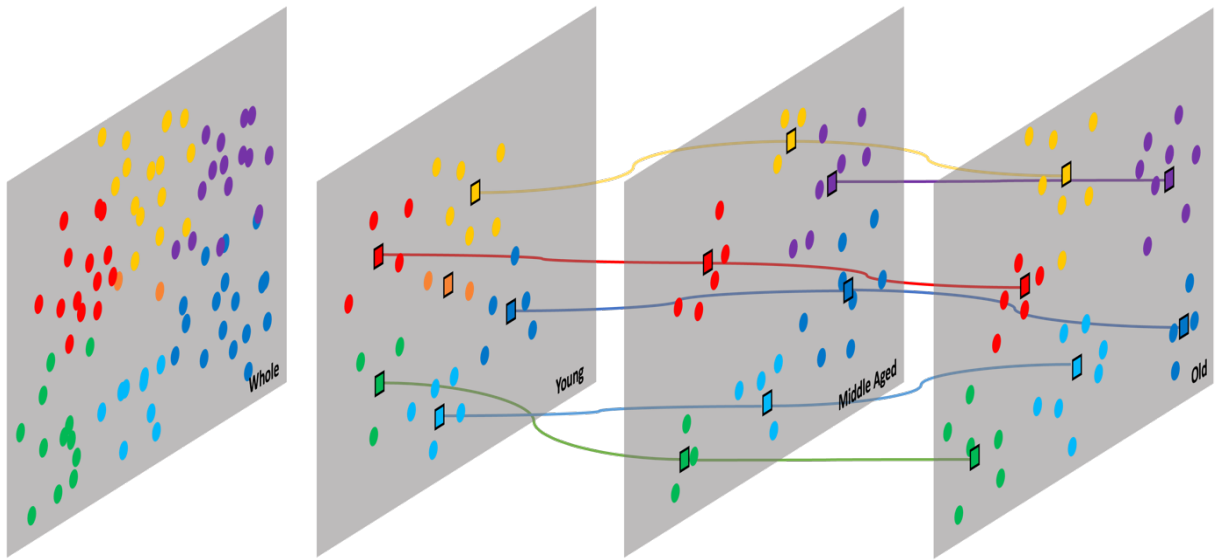

Figure 1: The figure illustrates age stratified clustering versus whole population clustering. In the figure, the left-hand slice represents all individuals (shown as circles) in the whole population, and the rest of the three slices represent the same individuals shows in their respective distinct age-groups; young, middle-aged and old. Each individual is part of a cluster set (represented by colour), and each square is the respective cluster centroid. Some cluster sets are singletons (only found in one age-group, e. g. orange in young population), others are non-singleton (found in varying numbers of consecutive age-groups, e.g., green in all age groups and blue in older age groups). We assume that there can be a potential shift in clusters found in different age-groups, so non-singleton clusters have to be similar but not necessarily identical. The grouping shown in the whole population is from the resulting cluster sets rather than from the whole population clustering (as evident from the overlaps among clusters).

Table 1: Prevalence of 40 long-term conditions across 12 age groups and in total. The proportion of patients for each group is shown in parentheses.

| Long-term condition full name (short name if applicable) | Prevalence (%-age) | | | | | | | | | | | | |
|---|---|---|---|---|---|---|---|---|---|---|---|---|---|
| | 1 (9.53) | 2 (10.40) | 3 (10.97) | 4 (12.16) | 5 (13.00) | 6 (11.51) | 7 (10.69) | 8 (9.10) | 9 (6.64) | 10 (4.04) | 11 (1.53) | 12 (0.44) | Total |
| Hypertension (HypTens) | 11.49 | 18.77 | 28.21 | 36.49 | 43.45 | 48.87 | 54.03 | 58.06 | 59.19 | 57.96 | 51.24 | 40.17 | 40.14 |
| Coronary Heart Disease (CHD) | 1.13 | 2.85 | 5.53 | 8.83 | 12.98 | 18.28 | 22.96 | 26.24 | 27.97 | 27.87 | 27.49 | 26.86 | 14.24 |
| Stroke or TIA (StrokeTIA) | 0.79 | 1.26 | 2.23 | 3.33 | 4.76 | 6.83 | 9.18 | 11.86 | 14.65 | 16.61 | 18.34 | 18.35 | 6.33 |
| Chronic kidney disease (CKD) | 0.48 | 0.67 | 1.05 | 1.81 | 3.07 | 5.59 | 8.94 | 13.36 | 16.62 | 18.36 | 16.78 | 14.59 | 5.83 |
| Peripheral Vascular Disease (PeriVascDis) | 1.22 | 1.56 | 1.87 | 2.51 | 3.45 | 4.77 | 5.55 | 6.45 | 7.01 | 6.81 | 5.34 | 4.16 | 3.81 |
| Atrial Fibrillation (AtrFib) | 0.42 | 0.56 | 0.92 | 1.47 | 2.51 | 3.97 | 6.02 | 8.41 | 11.17 | 13.57 | 15.3 | 14.91 | 4.16 |
| Heart Failure (HeartFail) | 0.39 | 0.7 | 1.06 | 1.48 | 2.13 | 3.19 | 4.6 | 6.02 | 7.87 | 9.88 | 12.77 | 14.79 | 3.27 |
| Diabetes | 5.84 | 7.44 | 9.4 | 11.26 | 12.6 | 15.16 | 17.2 | 16.83 | 14.77 | 12.4 | 10.35 | 8.79 | 12.16 |
| COPD | 3.53 | 4.44 | 5.32 | 6.8 | 8.58 | 10.98 | 12.18 | 13 | 12.69 | 10.62 | 8.28 | 7.31 | 8.5 |
| Asthma | 15.64 | 13.6 | 11.43 | 9.74 | 9.02 | 8.68 | 8.48 | 7.62 | 6.51 | 5.56 | 4.2 | 2.84 | 9.85 |
| Bronchiectasis | 0.14 | 0.18 | 0.24 | 0.43 | 0.56 | 0.6 | 0.75 | 0.64 | 0.66 | 0.53 | 0.45 | 0.28 | 0.47 |
| PainfulCondition (PainCond) | 14.66 | 16.14 | 18.16 | 20.23 | 22.26 | 23.47 | 23.76 | 22.01 | 20.31 | 19.49 | 16.53 | 13.39 | 20.11 |
| Depression (Depres) | 28.31 | 25.87 | 23.34 | 20.78 | 17.62 | 14.62 | 13.49 | 13.86 | 14.43 | 16.08 | 16.96 | 14.79 | 19.08 |
| Anxiety | 7.56 | 7.2 | 6.87 | 6.66 | 6.78 | 7.28 | 8.25 | 9.66 | 10.74 | 13.35 | 16.28 | 15.67 | 8.08 |
| Schizophrenia Bipolar (SchizBipol) | 2.68 | 2.49 | 2.04 | 1.75 | 1.54 | 1.26 | 1.17 | 1.22 | 1.05 | 1.18 | 0.97 | 0.6 | 1.67 |
| Dementia | 0.2 | 0.21 | 0.26 | 0.35 | 0.57 | 0.88 | 1.62 | 3.53 | 7 | 11.18 | 16.24 | 19.54 | 2.03 |
| Anorexia or Bulimia (Eating) | 1.27 | 0.87 | 0.6 | 0.35 | 0.23 | 0.19 | 0.22 | 0.28 | 0.24 | 0.3 | 0.37 | 0.32 | 0.46 |
| Learning Disability (LearnDisab) | 1.19 | 1.04 | 0.76 | 0.62 | 0.46 | 0.37 | 0.28 | 0.17 | 0.14 | 0.1 | 0.08 | 0.04 | 0.54 |
| Alcohol Problems (Alcohol) | 7.93 | 8.36 | 7.98 | 7.32 | 6.37 | 5.43 | 4.06 | 2.88 | 2 | 1.47 | 1 | 0.64 | 5.75 |
| Other Psychoactive Misuse (Substance) | 6.73 | 4.52 | 3.79 | 3.25 | 3.21 | 3.4 | 3.64 | 4.7 | 7.62 | 11.51 | 15.62 | 18.31 | 4.84 |
| Thyroid Disorders (Thyroid) | 8.65 | 9.36 | 10.21 | 10.75 | 11.1 | 11.63 | 12.34 | 13.21 | 14.03 | 14.71 | 15.62 | 15.99 | 11.36 |
| Inflammatory Arthritis (Arthritis) | 5.32 | 6.41 | 7.8 | 8.91 | 9.7 | 10.31 | 11.16 | 11.89 | 12.57 | 13.01 | 12.59 | 11.07 | 9.44 |
| Hearing Impairment (HearImpr) | 5.1 | 5.34 | 5.46 | 6.1 | 6.7 | 7.74 | 8.99 | 10.85 | 13.36 | 17.2 | 20.45 | 22.9 | 8.09 |
| Visual Impairment (VisualImpr) | 0.61 | 0.56 | 0.67 | 0.7 | 0.73 | 0.95 | 1.21 | 1.72 | 2.93 | 4.83 | 7.19 | 9.67 | 1.31 |
| Recent Cancer (Cancer) | 3.45 | 3.88 | 4.6 | 5.65 | 6.43 | 7.98 | 9.08 | 10.26 | 11.07 | 11.72 | 11.12 | 9.99 | 7.01 |
| Dyspepsia | 10.31 | 11.57 | 12.08 | 12.25 | 12.72 | 13.37 | 13.14 | 13.45 | 13.61 | 13.57 | 13.57 | 13.63 | 12.54 |
| Irritable Bowel Syndrome (IrritBSynd) | 10.54 | 10.03 | 9.03 | 7.79 | 6.65 | 5.61 | 5.1 | 4.43 | 3.83 | 3.68 | 2.74 | 2.36 | 6.9 |
| Constipation (Consti) | 2.07 | 2.21 | 2.51 | 2.93 | 3.78 | 4.97 | 7.13 | 9.89 | 14.14 | 19.09 | 25.88 | 30.26 | 6.02 |
| Diverticular Disease (Diverticlr) | 0.68 | 1.36 | 2.37 | 3.63 | 4.98 | 6.65 | 8.63 | 11.03 | 12.47 | 13.68 | 13.26 | 12.99 | 5.89 |
| Inflammatory Bowel Disease (InflmBDis) | 1.76 | 1.66 | 1.51 | 1.5 | 1.37 | 1.24 | 1.18 | 1.13 | 1 | 0.85 | 0.63 | 0.44 | 1.35 |
| Chronic Sinusitis (Sinus) | 1.66 | 1.85 | 1.76 | 1.66 | 1.42 | 1.1 | 0.99 | 0.79 | 0.63 | 0.59 | 0.38 | 0.48 | 1.31 |
| Viral Hepatitis (Viral Hepat) | 0.47 | 0.26 | 0.18 | 0.09 | 0.04 | 0.02 | 0.03 | 0.01 | 0.02 | 0.01 | 0.01 | 0 | 0.12 |
| Chronic Liver Disease | 0.29 | 0.47 | 0.58 | 0.62 | 0.63 | 0.56 | 0.36 | 0.25 | 0.15 | 0.07 | 0.05 | 0.04 | 0.44 |
| Prostate Disorders (Prostat) | 0.56 | 0.82 | 1.13 | 1.72 | 2.41 | 3.27 | 4.14 | 4.7 | 5.09 | 4.5 | 3.31 | 2.4 | 2.61 |
| Glaucoma | 0.39 | 0.7 | 1.17 | 1.59 | 2 | 2.82 | 3.61 | 4.94 | 6.17 | 7.7 | 8.91 | 8.39 | 2.75 |
| Epilepsy | 2.29 | 2.19 | 1.88 | 1.62 | 1.49 | 1.34 | 1.24 | 1.17 | 1.01 | 0.95 | 0.88 | 0.44 | 1.56 |
| Migraine | 1.89 | 2.19 | 2.52 | 2.02 | 1.46 | 0.94 | 0.5 | 0.32 | 0.21 | 0.17 | 0.09 | 0.12 | 1.33 |
| Parkinsons | 0.02 | 0.04 | 0.09 | 0.17 | 0.26 | 0.46 | 0.69 | 1.05 | 1.43 | 1.61 | 1.38 | 1.4 | 0.48 |
| Multiple Sclerosis (MultScleros) | 0.83 | 0.91 | 0.92 | 0.76 | 0.6 | 0.44 | 0.31 | 0.27 | 0.18 | 0.13 | 0.09 | 0.04 | 0.57 |
| Psoriasis or Eczema (PsoriaEczma) | 1.71 | 1.6 | 1.39 | 1.28 | 1.22 | 1.27 | 1.29 | 1.17 | 1.15 | 1.08 | 0.93 | 0.64 | 1.32 |

TIA: transient ischaemic attack; COPD: chronic obstructive pulmonary disease

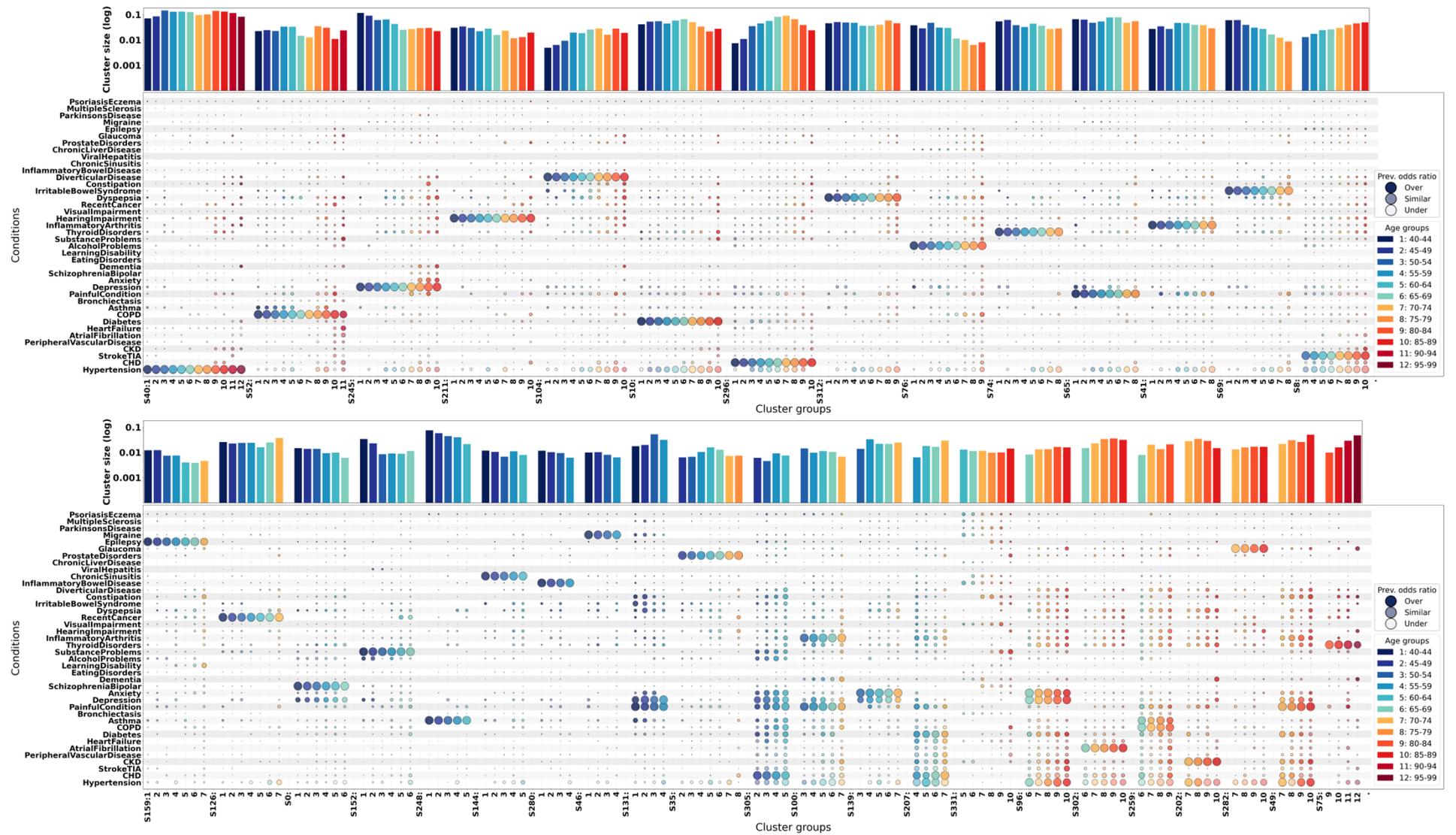

Figure 2: Non-singleton cluster sets. Sets with #clusters ≥ 8 (up) and sets with #clusters ∈ {4, 5, 6, 7} (down) are displayed.

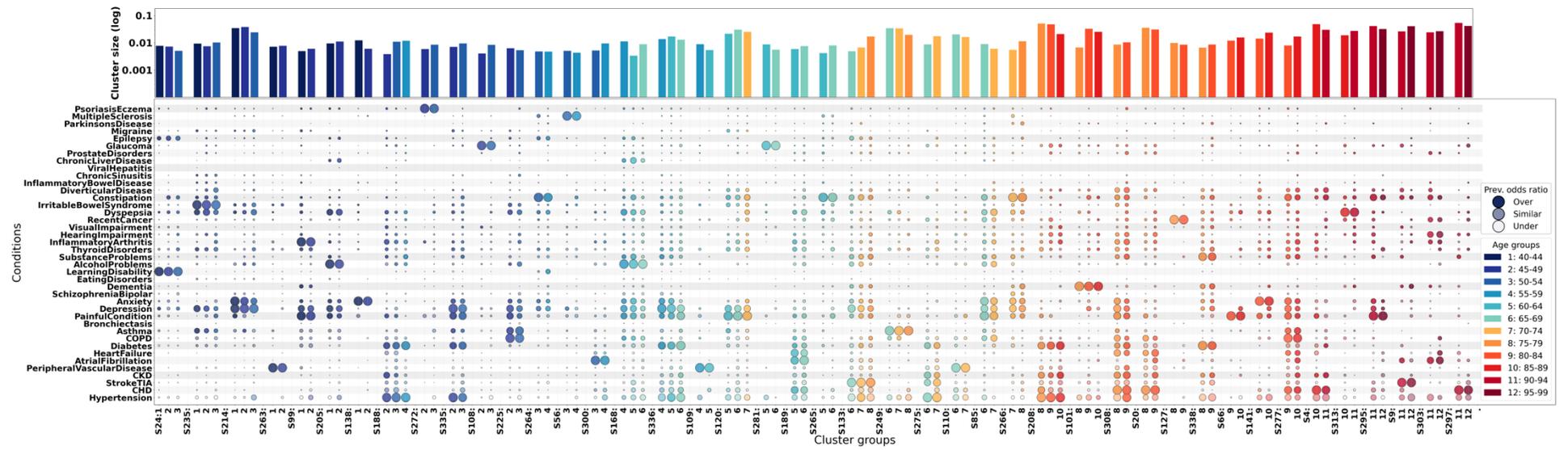

Figure 3: Non-singleton cluster sets. Sets with #clusters ∈ {2, 3} are displayed.

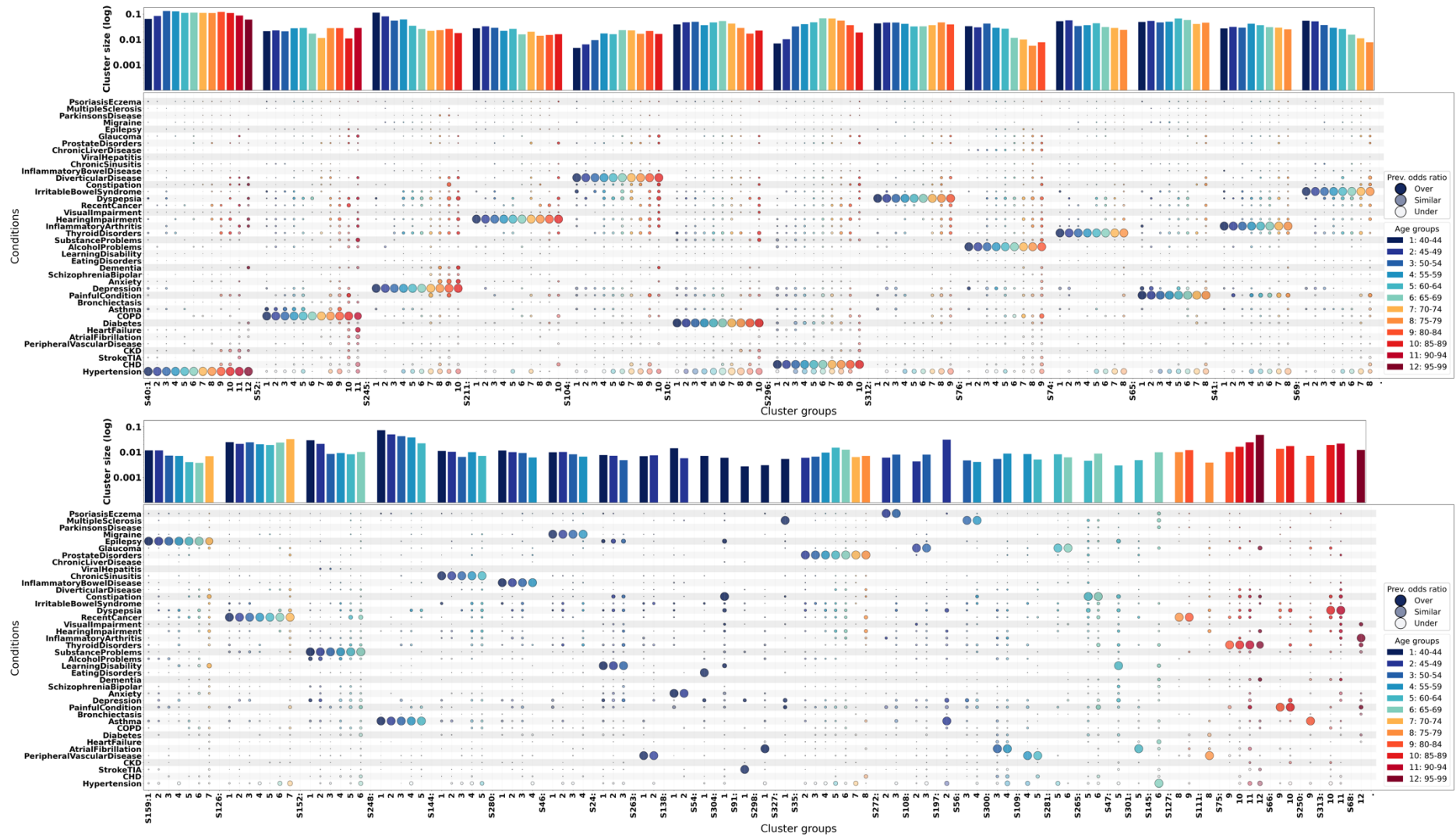

Figure 4: Single LTC dominant cluster sets.

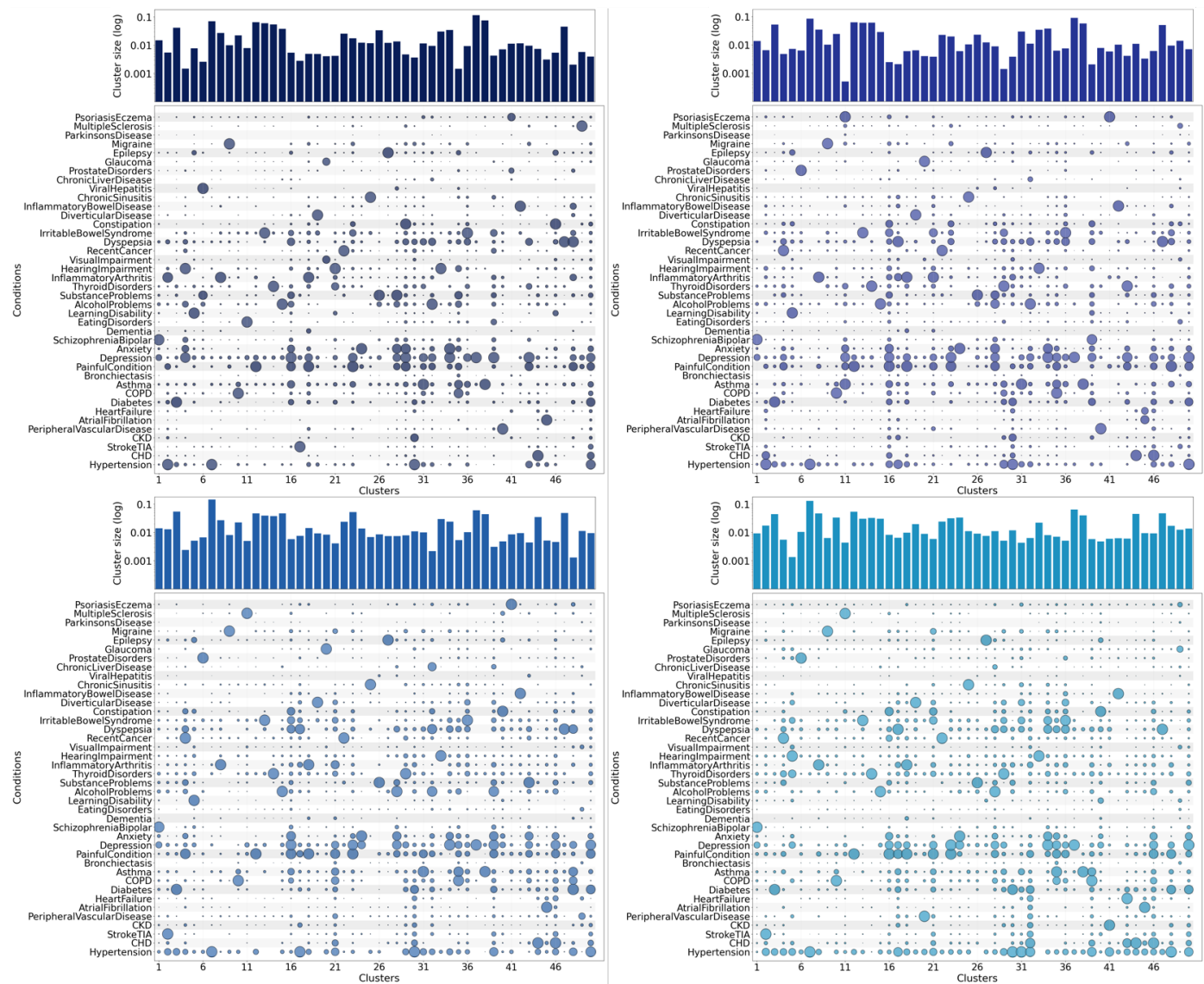

Supplement Figure 1: Cluster prevalence for best LCA runs (K = 50) on age-groups (from left) 1 : 40 – 44, 2 : 45 – 49, 3 : 50 – 54, and 4 : 55 – 59.

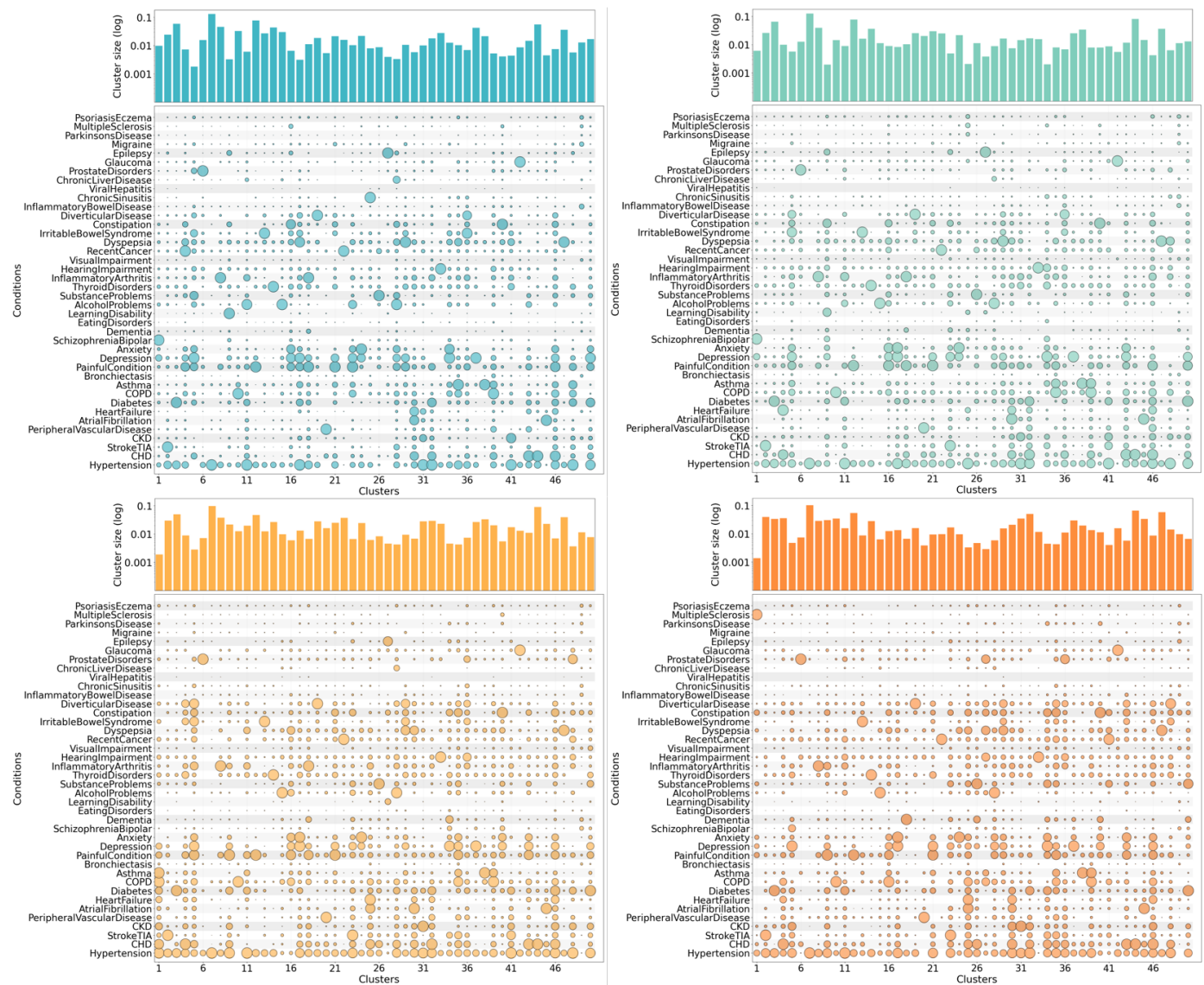

Supplement Figure 2: Cluster prevalence for best LCA runs (K = 50) on age-groups (from left) 5 : 60 – 64, 6 : 65 – 69, 7 : 70 – 74, and 8 : 75 – 79.

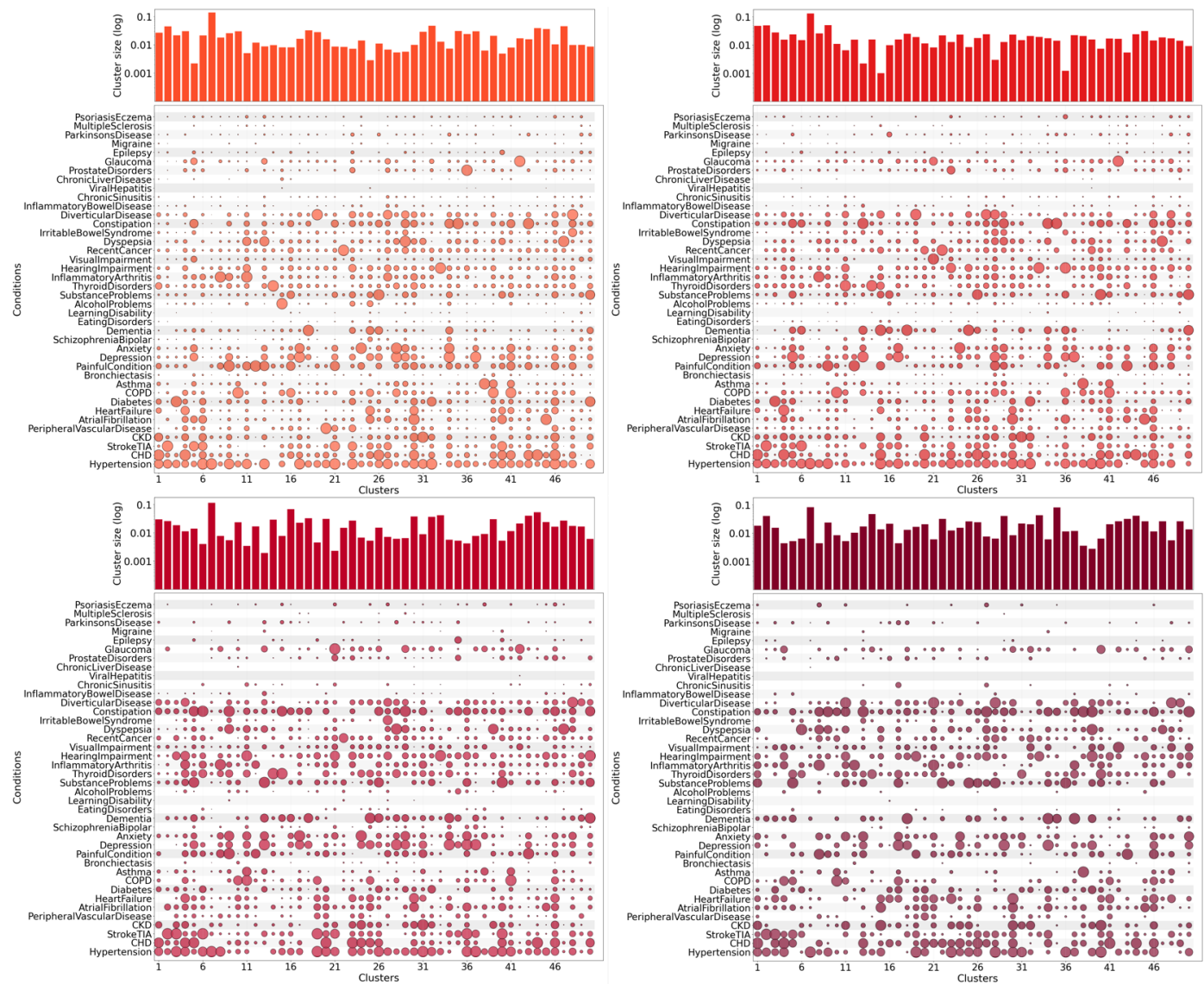

Supplement Figure 3: Cluster prevalence for best LCA runs (K = 50) on age-groups (from left) 9 : 80 – 84, 10 : 85 – 89, 11 : 90 – 94, and 12 : 95 – 99.

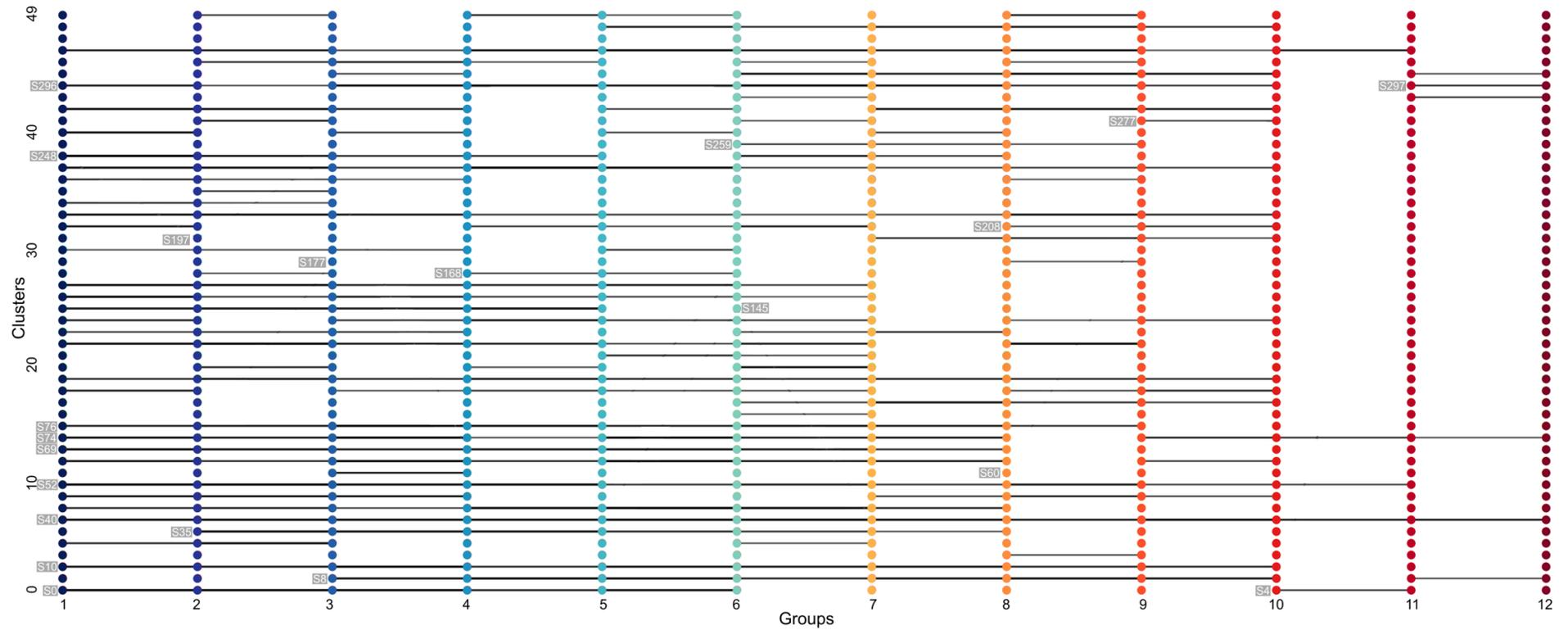

Supplement Figure 4: Similarity network showing similarities between clusters of different age-groups starting from (left) the age-group 1 (40 – 44 years) to 12 (95 – 99 years) for best LCA runs (K = 50). Each column represents K = 50 cluster sets observed in a specific age-group. Each node represents a cluster set. Singleton cluster sets have no edge. Non-singleton cluster sets have edges to the next oldest age-group. The thicker the edge, the higher the similarity between two clusters. Clusters that are highly similar (connected by strong edges) are merged and considered as a non-singleton cluster set, and visualized in separate figures. To achieve a clearer visualization of the similarity network in this figure, we show only the edges where $S(\theta_i, \theta_j) \geq 0.7$.

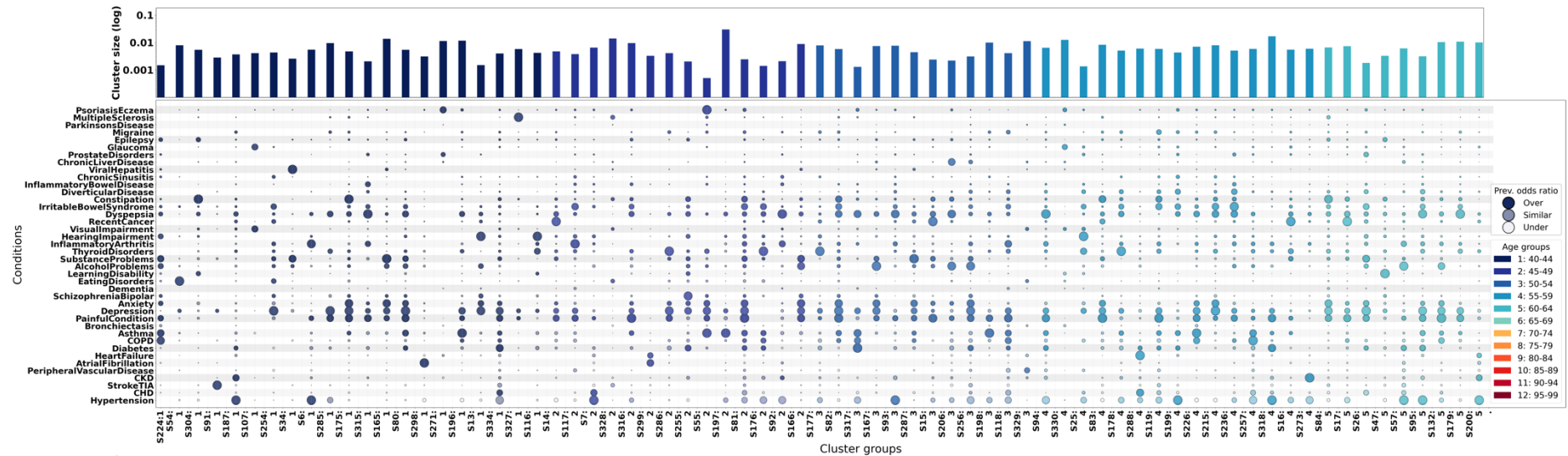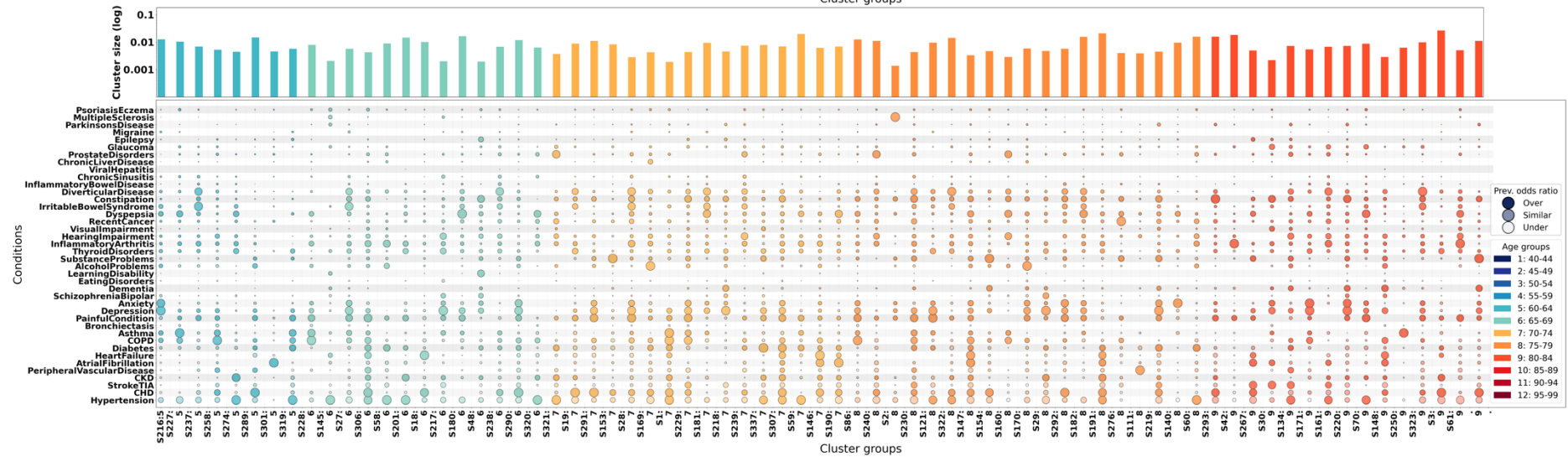

Supplement Figure 5:: Singleton cluster sets (part 1)

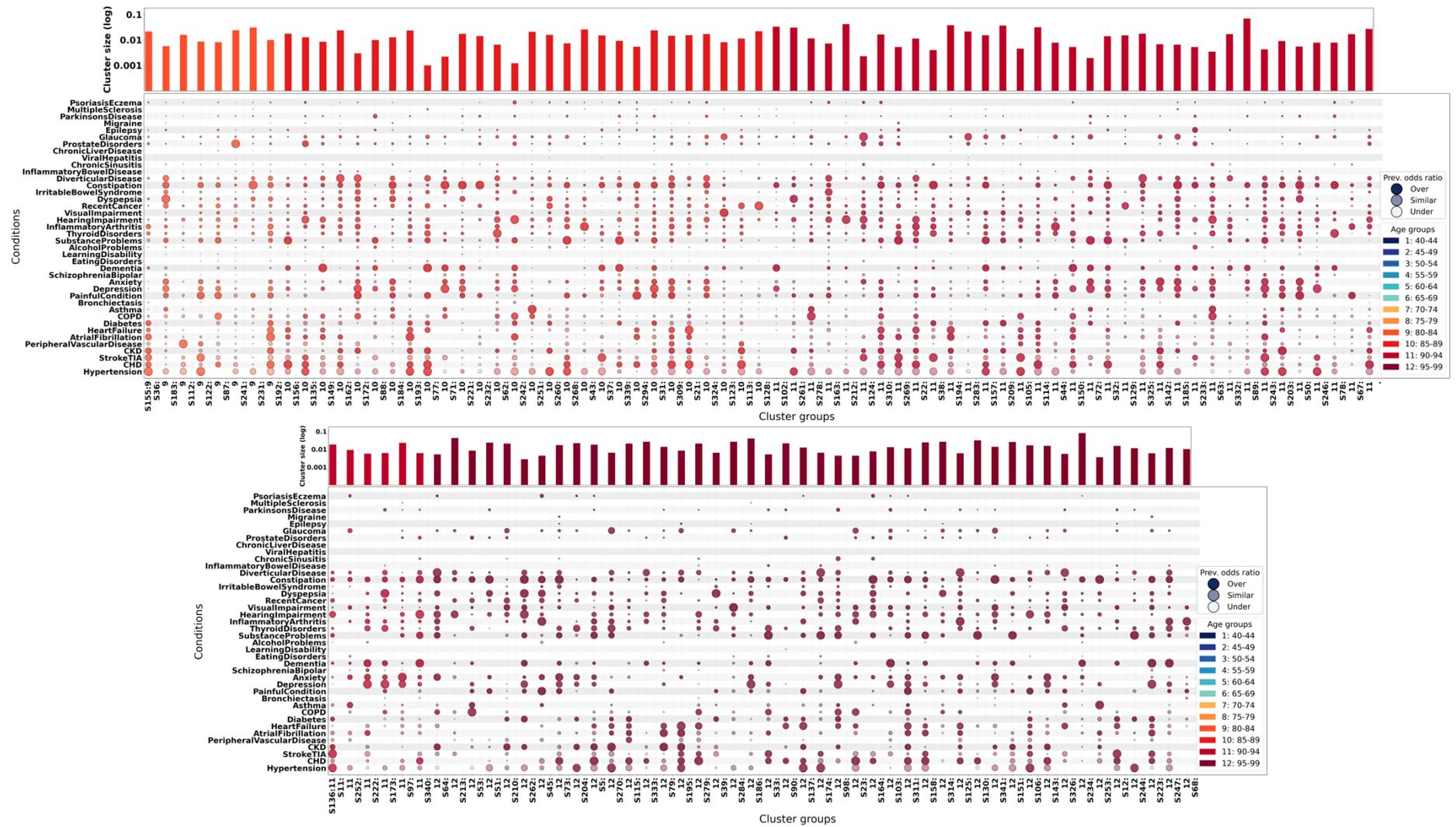

Supplement Figure 6: Singleton cluster sets (part 2)

Supplement Table 1 First cluster (i.e., the representative) of every cluster set with no moderate conditions (i.e., prevalence is between 0.3 and 0.7) and more than 30 conditions of low prevalence (i.e., prevalence is below 0.3).

| Distinct LTC (prevalence > 0.9) | Moderate LTCs | Cluster size in age groups (% of age group) | | | | | | | | | | | | Cluster Set |
|---|---|---|---|---|---|---|---|---|---|---|---|---|---|---|
| | | 1 | 2 | 3 | 4 | 5 | 6 | 7 | 8 | 9 | 10 | 11 | 12 | |
| HypTens | - | 6.46 | 8.35 | 13.10 | 12.84 | 11.14 | 11.42 | 11.01 | 10.88 | 12.12 | 10.91 | 8.73 | 6.05 | S40 |
| COPD | - | 2.11 | 2.28 | 2.10 | 2.77 | 2.84 | 1.68 | 1.14 | 2.79 | 2.78 | 1.09 | 2.83 | | S52 |
| Depres | - | 11.36 | 8.08 | 5.47 | 6.09 | 3.46 | 2.58 | 2.17 | 2.33 | 2.62 | 1.79 | | | S245 |
| HearImpr | - | 2.75 | 3.26 | 2.88 | 2.18 | 2.61 | 1.59 | 2.01 | 1.40 | 1.49 | 1.61 | | | S211 |
| Diverticlr | - | 0.46 | 0.64 | 0.94 | 1.72 | 1.60 | 2.32 | 2.27 | 1.67 | 2.16 | 1.64 | | | S104 |
| Diabetes | - | 3.88 | 4.75 | 4.91 | 3.64 | 4.71 | 5.26 | 4.24 | 2.83 | 1.69 | 2.24 | | | S10 |
| CHD | - | 0.70 | 1.02 | 3.24 | 4.01 | 4.76 | 6.76 | 6.70 | 5.49 | 3.62 | 1.88 | | | S296 |
| Dyspepsia | - | 4.29 | 4.65 | 4.63 | 4.12 | 3.23 | 3.28 | 3.66 | 4.67 | 3.89 | | | | S312 |
| Alcohol | - | 3.30 | 2.98 | 4.20 | 2.86 | 2.64 | 1.15 | 0.99 | 0.57 | 0.77 | | | | S76 |
| Thyroid | - | 5.23 | 5.67 | 3.37 | 3.64 | 4.32 | 3.07 | 2.84 | 2.39 | | | | | S74 |
| PainCond | - | 4.90 | 5.35 | 4.66 | 4.98 | 6.68 | 5.81 | 4.06 | 4.58 | | | | | S65 |
| Arthritis | - | 2.74 | 3.10 | 2.91 | 4.17 | 3.66 | 3.03 | 2.92 | 2.51 | | | | | S41 |
| IrritBSynd | - | 5.39 | 5.16 | 3.69 | 2.87 | 2.59 | 1.53 | 1.10 | 0.78 | | | | | S69 |
| Epilepsy | - | 1.17 | 1.16 | 0.72 | 0.70 | 0.40 | 0.37 | 0.69 | | | | | | S159 |
| Cancer | - | 2.49 | 2.13 | 2.46 | 2.03 | 1.89 | 2.38 | 3.29 | | | | | | S126 |
| Substance | - | 2.95 | 2.12 | 0.85 | 0.92 | 0.82 | 1.01 | | | | | | | S152 |
| Asthma | - | 7.36 | 4.99 | 4.26 | 3.80 | 2.22 | | | | | | | | S248 |
| Sinus | - | 1.11 | 1.02 | 0.65 | 1.00 | 0.71 | | | | | | | | S144 |
| InflmBDis | - | 1.15 | 1.00 | 0.91 | 0.61 | | | | | | | | | S280 |
| Migraine | - | 0.98 | 1.01 | 0.81 | 0.66 | | | | | | | | | S46 |
| LearnDisab | - | 0.76 | 0.72 | 0.48 | | | | | | | | | | S24 |
| PeriVascDis | - | 0.69 | 0.74 | | | | | | | | | | | S263 |
| Anxiety | - | 1.42 | 0.57 | | | | | | | | | | | S138 |
| Eating | - | 0.71 | | | | | | | | | | | | S54 |
| Consti | - | 0.60 | | | | | | | | | | | | S304 |

| | | | | | | | | | | | | |
|---|---|---|---|---|---|---|---|---|---|---|---|---|
| StrokeTIA | - | 0.27 | | | | | | | | | | S91 |
| AtrFib | - | 0.30 | | | | | | | | | | S298 |
| MultScleros | - | 0.54 | | | | | | | | | | S327 |
| Prostat | - | | 0.60 | 0.66 | 0.96 | 1.49 | 1.25 | 0.63 | 0.71 | | | S35 |
| PsoriaEczma | - | | 0.60 | 0.79 | | | | | | | | S272 |
| Glaucoma | - | | 0.42 | 0.79 | | | | | | | | S108 |
| LTC (prevalence > 0.7) | | 1 | 2 | 3 | 4 | 5 | 6 | 7 | 8 | 9 | 10 | 11 | 12 |
| Depres Anxiety | - | 2.56 | 5.39 | 4.18 | | | | | | | | S214 |
| PainCond Arthritis | - | | | 1.16 | 0.90 | 1.11 | 1.12 | 0.88 | | | | S100 |
| HypTens Depres Anxiety | - | | | | | 0.75 | 1.47 | 1.19 | 1.57 | 1.73 | | S96 |
| Cancer | - | | | | | | | | | | 1.34 | S129 |
| Arthritis | - | | | | | | | | | | 1.22 | S68 |

Supplement Table 2: First cluster (i.e., the representative) of every cluster set with one moderate condition (i.e., prevalence is between 0.3 and 0.7) and more than 30 conditions of low prevalence (i.e., prevalence is below 0.3).

| LTC (prevalence > 0.7) | Moderate LTC (Prev. 0.5 – 0.7) | Cluster size in age groups (% of age group) | | | | | | | | | | | | Cluster Set |
|---|---|---|---|---|---|---|---|---|---|---|---|---|---|---|
| | | 1 | 2 | 3 | 4 | 5 | 6 | 7 | 8 | 9 | 10 | 11 | 12 | |
| PainCond Arthritis | Depres | 0.59 | 0.51 | | | | | | | | | | | S99 |
| HypTens | CKD | 0.83 | | | | | | | | | | | | S187 |
| ViralHepat | Substance | 0.26 | | | | | | | | | | | | S34 |
| HearImpr | Thyroid | 0.81 | | | | | | | | | | | | S116 |
| Depres Thyroid | PainCond | | 0.53 | | | | | | | | | | | S286 |
| Anxiety | Depres | | | 1.18 | 2.71 | 1.75 | 2.08 | 1.89 | | | | | | S139 |
| CKD | HypTens | | | | 0.53 | | | | | | | | | S273 |
| Arthritis | HypTens | | | | | | | | | | 1.92 | | | S43 |
| VisualImpr | Glaucoma | | | | | | | | | | 0.82 | | | S123 |
| HypTens COPD | Asthma | | | | | | | | | | | 1.15 | | S278 |
| PainCond | Arthritis | | | | | | | | | | | 1.43 | | S67 |
| Substance | CHD | | | | | | | | | | | | 2.48 | S130 |
| LTC (prevalence > 0.7) | Moderate LTC (Prev. 0.3 – 0.5) | 1 | 2 | 3 | 4 | 5 | 6 | 7 | 8 | 9 | 10 | 11 | 12 | |
| SchizBipol | Depres | 1.34 | 1.24 | 1.28 | 0.83 | 0.89 | 0.53 | | | | | | | S0 |
| HypTens Arthritis | PainCond | 0.69 | | | | | | | | | | | | S6 |
| Depres PainCond Anxiety Consti | Dyspepsia | 0.58 | | | | | | | | | | | | S175 |
| StrokeTIA | HypTens | | | 1.16 | 1.43 | 2.06 | 2.03 | 2.4 | 3.06 | 3.36 | 3.5 | | | S8 |
| Consti | Depres | | | 0.55 | 0.69 | | | | | | | | | S264 |
| HypTens Dyspepsia | PainCond | | | 1.33 | | | | | | | | | | S93 |
| HypTens PainCond Diabetes | Depres | | | | 2.76 | | | | | | | | | S318 |
| HypTens | CKD | | | | | 3.9 | | | | | | | | S200 |
| AtrFib | HypTens | | | | | | 1.14 | 1.84 | 2.59 | 2.81 | 2.48 | | | S302 |
| Asthma | HypTens | | | | | | 3.13 | 2.77 | 1.83 | | | | | S249 |
| PeriVascDis | HypTens | | | | | | 1.65 | 1.25 | | | | | | S110 |
| CHD HeartFail | HypTens | | | | | | | 0.89 | | | | | | S18 |
| Glaucoma | HypTens | | | | | | | 1.17 | 1.28 | 1.41 | 1.43 | | | S282 |

| | | | | | | | | | | | | | |
|---|---|---|---|---|---|---|---|---|---|---|---|---|---|
| Dementia | HypTens | | | | | | | | 0.62 | 2.52 | 1.86 | | S101 |
| MultScleros | HypTens | | | | | | | | 0.12 | | | | S2 |
| HypTens Depres PainCond | Thyroid | | | | | | | | 1.31 | | | | S121 |
| Anxiety | Dyspepsia | | | | | | | | 1.12 | | | | S140 |
| Arthritis | PainCond | | | | | | | | | 1.79 | | | S42 |
| Prostat | HypTens | | | | | | | | | 1.97 | | | S241 |
| Cancer | HypTens | | | | | | | | | | 1.8 | | S128 |
| CHD | HypTens | | | | | | | | | | | 3.65 | 3.46 | S297 |
| HearImpr | HypTens | | | | | | | | | | | 3.3 | | S212 |
| VisualImpr | HypTens | | | | | | | | | | | 2.11 | | S284 |
| HypTens Substance | Diabetes | | | | | | | | | | | | 1.05 | S244 |

Supplement Table 3: First cluster (i.e., the representative) of every cluster set with two moderate conditions (i.e., prevalence is between 0.3 and 0.7) and more than 30 conditions of low prevalence (i.e., prevalence is below 0.3).

| LTC (prevalence > 0.7) | Moderate LTCs | Cluster size in age groups (% of age group) | | | | | | | | | | | | Cluster Set |
|---|---|---|---|---|---|---|---|---|---|---|---|---|---|---|
| | | 1 | 2 | 3 | 4 | 5 | 6 | 7 | 8 | 9 | 10 | 11 | 12 | |
| Depres PainCond | Asthma IrritBSynd | 2.56 | 2.4 | 4.48 | 2.32 | | | | | | | | | S131 |
| Depres | IrritBSynd Eating | 1.47 | | | | | | | | | | | | S254 |
| Depres HearImpr | Anxiety Alcohol | 0.21 | | | | | | | | | | | | S13 |
| HypTens Depres | PainCond Diabetes | | 0.83 | 1.23 | | | | | | | | | | S335 |
| Dyspepsia | HypTens Arthritis | | 0.49 | | | | | | | | | | | S92 |
| Depres PainCond Anxiety | Alcohol Substance | | 1.2 | | | | | | | | | | | S166 |
| Thyroid | HypTens Depres | | | | 1.67 | | | | | | | | | S177 |
| PainCond Cancer | Dyspepsia Substance | | | | 0.27 | | | | | | | | | S15 |
| Asthma | HypTens PainCond | | | | 1.37 | | | | | | | | | S198 |
| HeartFail | HypTens CHD | | | | | 0.59 | | | | | | | | S288 |
| HypTens COPD | Asthma CHD | | | | | 0.76 | | | | | | | | S257 |
| Cancer | HypTens Thyroid | | | | | 1 | | | | | | | | S16 |
| PainCond | HypTens Depres | | | | | 2.49 | 3.56 | 2.83 | | | | | | S120 |
| AtrFib HeartFail | HypTens CHD | | | | | 0.91 | 0.83 | | | | | | | S189 |
| Depres Consti | PainCond Anxiety | | | | | 0.96 | | | | | | | | S84 |
| HypTens Alcohol | Diabetes Arthritis | | | | | 1.05 | | | | | | | | S57 |
| Depres Anxiety | HypTens Dyspepsia | | | | | 1.88 | | | | | | | | S216 |
| HypTens CKD | Dyspepsia Thyroid | | | | | 0.99 | | | | | | | | S274 |
| CHD | HypTens PainCond | | | | | 1.8 | | | | | | | | S289 |
| Depres StrokeTIA | HypTens Consti | | | | | | 0.65 | 0.82 | 1.67 | | | | | S133 |
| Substance | HypTens CHD | | | | | | | 0.74 | | | | | | S153 |
| Diabetes Substance | HypTens Dementia | | | | | | | | 0.75 | 1.12 | | | | S338 |
| CHD Substance | HypTens Dementia | | | | | | | | 0.6 | | | | | S154 |
| Anxiety | HypTens PainCond | | | | | | | | | 1.25 | 2.1 | | | S141 |
| CHD Consti | HypTens PainCond | | | | | | | | | 1.54 | | | | S293 |
| HypTens PainCond Dyspepsia | Depres Anxiety | | | | | | | | | 1.69 | | | | S70 |

| | | | | | | | | | | | |
|---|---|---|---|---|---|---|---|---|---|---|---|
| HypTens CHD | Thyroid CKD | | | | | | | 3.27 | | | S3 |
| PeriVascDis | HypTens CHD | | | | | | | 1.16 | | | S112 |
| HypTens StrokeTIA | Consti Dementia | | | | | | | | 2.03 | | S37 |
| Substance Dementia | CHD Consti | | | | | | | | 1.21 | | S339 |
| StrokeTIA | HypTens AtrFib | | | | | | | | 2.13 | 2.79 | S9 |
| Dementia | Depres Consti | | | | | | | | 2.94 | | S102 |
| HypTens Dyspepsia | HearImpr CKD | | | | | | | | 3.3 | | S261 |
| PainCond Dyspepsia Anxiety Consti | Depres Thyroid | | | | | | | | 0.57 | | S50 |
| Thyroid Consti | HypTens Dementia | | | | | | | | 0.73 | | S78 |
| Asthma Consti | CHD AtrFib | | | | | | | | | 0.32 | S253 |

Supplement Table 4: First cluster (i.e., the representative) of every cluster set with three moderate conditions (i.e., prevalence is between 0.3 and 0.7) and more than 30 conditions of low prevalence (i.e., prevalence is below 0.3).

| LTC (prevalence > 0.7) | Moderate LTCs | Cluster size in age groups (% of age group) | | | | | | | | | | | | Cluster Set |
|---|---|---|---|---|---|---|---|---|---|---|---|---|---|---|
| | | 1 | 2 | 3 | 4 | 5 | 6 | 7 | 8 | 9 | 10 | 11 | 12 | |
| IrritBSynd | Depres Asthma Dyspepsia | 1.47 | 1.28 | 1.40 | | | | | | | | | | S235 |
| Alcohol | Depres PainCond Dyspepsia | 1.25 | 1.38 | | | | | | | | | | | S205 |
| Anxiety Substance | Depres PainCond Alcohol | 1.50 | | | | | | | | | | | | S165 |
| Asthma | Depres PainCond Dyspepsia | 2.25 | | | | | | | | | | | | S196 |
| CHD | Depres PainCond Diabetes | | 0.63 | 0.51 | 1.05 | 0.89 | | | | | | | | S305 |
| HypTens | Diabetes Arthritis CKD | | 0.75 | 2.10 | 1.68 | | | | | | | | | S188 |
| Cancer | Depres PainCond Dyspepsia | | 0.58 | | | | | | | | | | | S14 |
| Arthritis | Depres PainCond IrritBSynd | | 0.61 | | | | | | | | | | | S117 |
| Depres Asthma PsoriaEczma | PainCond Thyroid Anxiety | | 0.05 | | | | | | | | | | | S55 |
| Substance | Depres PainCond Anxiety | | | 0.51 | | | | | | | | | | S287 |
| Dyspepsia Alcohol | HypTens PainCond ChronicLivr | | | 0.36 | | | | | | | | | | S206 |
| Depres | Dyspepsia Anxiety IrritBSynd | | | | 1.27 | | | | | | | | | S215 |
| Depres PainCond | COPD Anxiety Alcohol | | | | | 1.32 | | | | | | | | S132 |
| IrritBSynd Diverticlr | HypTens PainCond Arthritis | | | | | 1.19 | | | | | | | | S237 |
| Asthma COPD | HypTens PainCond CHD | | | | | | 1.03 | 2.58 | 1.72 | 2.81 | | | | S259 |
| Dyspepsia | HypTens Depres PainCond | | | | | | 1.96 | | | | | | | S180 |
| Diverticlr | HypTens CHD IrritBSynd | | | | | | 1.38 | | | | | | | S238 |
| Consti | Depres PainCond Anxiety | | | | | | | 0.60 | 1.50 | | | | | S266 |
| PainCond CHD | Depres COPD Anxiety | | | | | | | 1.18 | | | | | | S291 |
| HypTens Diverticlr | CHD Thyroid IrritBSynd | | | | | | | | | 1.97 | | | | S322 |
| Depres Anxiety | Dyspepsia Thyroid COPD | | | | | | | | | 0.56 | | | | S171 |
| HypTens Arthritis | PainCond Dyspepsia HearImpr | | | | | | | | | 0.59 | | | | S61 |
| Thyroid | HypTens HearImpr Diverticlr | | | | | | | | | | 0.81 | | | S62 |
| HypTens CHD | Cancer CKD PeriVascDis | | | | | | | | | | 1.47 | | | S113 |
| AtrFib HeartFail | HypTens CHD CKD | | | | | | | | | | | 4.47 | | S194 |

| | | | | | | | | | | | | | |
|---|---|---|---|---|---|---|---|---|---|---|---|---|---|
| HearImpr Consti Dementia | CHD Substance Diverticlr | | | | | | | | | | 0.67 | | S340 |
| COPD | Asthma Cancer Consti | | | | | | | | | | | 0.82 | S53 |
| PainCond Dyspepsia Arthritis Consti | CHD Anxiety Cancer | | | | | | | | | | | 0.47 | S45 |
| IrritBSynd | Depres Asthma Dyspepsia | 1.47 | 1.28 | 1.40 | | | | | | | | | S235 |
| Alcohol | Depres PainCond Dyspepsia | 1.25 | 1.38 | | | | | | | | | | S205 |
| Anxiety Substance | Depres PainCond Alcohol | 1.50 | | | | | | | | | | | S165 |
| Asthma | Depres PainCond Dyspepsia | 2.25 | | | | | | | | | | | S196 |
| CHD | Depres PainCond Diabetes | | 0.63 | 0.51 | 1.05 | 0.89 | | | | | | | S305 |
| HypTens | Diabetes Arthritis CKD | | 0.75 | 2.10 | 1.68 | | | | | | | | S188 |
| Cancer | Depres PainCond Dyspepsia | | 0.58 | | | | | | | | | | S14 |

Supplement Table 5: First cluster (i.e., the representative) of every cluster set with four moderate conditions (i.e., prevalence is between 0.3 and 0.7) and more than 30 conditions of low prevalence (i.e., prevalence is below 0.3).

| LTC (prevalence > 0.7) | Moderate LTCs | Cluster size in age groups (% of age group) | | | | | | | | | | | | Cluster Set |
|---|---|---|---|---|---|---|---|---|---|---|---|---|---|---|
| | | 1 | 2 | 3 | 4 | 5 | 6 | 7 | 8 | 9 | 10 | 11 | 12 | |
| Depres | HypTens PainCond Dyspepsia Thyroid | 1.75 | | | | | | | | | | | | S285 |
| Dyspepsia | Depres PainCond Arthritis InflmBDis | 0.45 | | | | | | | | | | | | S315 |
| HypTens | Depres PainCond CHD Diabetes | 0.54 | | | | | | | | | | | | S334 |
| COPD | Depres PainCond Asthma Dyspepsia | | 0.60 | 0.74 | | | | | | | | | | S225 |
| PainCond | Depres Asthma Dyspepsia IrritBSynd | | 1.60 | | | | | | | | | | | S316 |
| SchizBipol | Depres Dyspepsia Diabetes Consti | | 0.35 | | | | | | | | | | | S255 |
| Depres PainCond Anxiety | Asthma Dyspepsia IrritBSynd Consti | | | 0.82 | | | | | | | | | | S82 |
| Asthma Diabetes | HypTens Depres Dyspepsia COPD | | | 0.17 | | | | | | | | | | S317 |
| Depres Alcohol | PainCond Dyspepsia Anxiety Substance | | | 0.73 | | | | | | | | | | S167 |
| Alcohol | Depres PainCond Dyspepsia Anxiety | | | | 1.23 | 0.48 | 1.02 | | | | | | | S168 |
| PainCond Dyspepsia | HypTens Depres Asthma Diabetes | | | | 0.80 | | | | | | | | | S94 |
| HearImpr | HypTens Depres Thyroid IrritBSynd | | | | 0.27 | | | | | | | | | S25 |
| Thyroid | HypTens Depres Dyspepsia IrritBSynd | | | | 0.90 | | | | | | | | | S178 |
| IrritBSynd | Depres PainCond Dyspepsia Diverticlr | | | | 0.84 | | | | | | | | | S236 |
| Cancer | HypTens Depres PainCond Dyspepsia | | | | | 0.79 | | | | | | | | S17 |
| HypTens Depres Dyspepsia | PainCond Asthma Diabetes IrritBSynd | | | | | 0.37 | | | | | | | | S95 |
| HypTens Prostat | Diabetes HearImpr Cancer CKD | | | | | | | 0.76 | | | | | | S321 |
| CHD AtrFib HeartFail | HypTens Arthritis COPD CKD | | | | | | | 0.67 | | | | | | S146 |
| Depres Anxiety Consti | PainCond CHD Substance Dementia | | | | | | | | | 0.90 | | | | S220 |
| HypTens PainCond StrokeTIA | Depres Anxiety Consti PeriVascDis | | | | | | | | | 1.12 | | | | S122 |
| HypTens HearImpr | CHD COPD StrokeTIA Prostat | | | | | | | | | | 1.90 | | | S135 |
| Depres Consti | COPD Anxiety StrokeTIA Dementia | | | | | | | | | | 0.22 | | | S71 |
| CHD HeartFail | HypTens HearImpr CKD AtrFib | | | | | | | | | | 1.54 | | | S21 |
| HypTens Consti StrokeTIA | Diabetes Thyroid Arthritis HearImpr | | | | | | | | | | | 0.40 | | S38 |
| Depres Substance | Dyspepsia Anxiety Glaucoma Dementia | | | | | | | | | | | 0.19 | | S72 |

| | | | | | | | | | | | | |
|---|---|---|---|---|---|---|---|---|---|---|---|---|
| CHD Anxiety | HypTens Depres Thyroid CKD | | | | | | | | | | 0.63 | S142 |
| CKD AtrFib | HypTens Dyspepsia Cancer HeartFail | | | | | | | | | | 1.50 | S79 |
| HypTens CHD CKD AtrFib HeartFail | Consti StrokeTIA Diverticlr Glaucoma | | | | | | | | | | 0.80 | S195 |
| Dementia | Depres HearImpr Consti StrokeTIA | | | | | | | | | | 1.84 | S103 |

Supplement Table 6: First cluster (i.e., the representative) of every cluster set with five moderate conditions (i.e., prevalence is between 0.3 and 0.7) and more than 30 conditions of low prevalence (i.e., prevalence is below 0.3).

| LTC (prevalence > 0.7) | Moderate LTCs | Cluster size in age groups (% of age group) | | | | | | | | | | | | Cluster Set |
|---|---|---|---|---|---|---|---|---|---|---|---|---|---|---|
| | | 1 | 2 | 3 | 4 | 5 | 6 | 7 | 8 | 9 | 10 | 11 | 12 | |
| COPD | PainCond Asthma HearImpr Alcohol Substance | 0.24 | | | | | | | | | | | | S224 |
| Depres PainCond | Asthma Dyspepsia Anxiety Alcohol Substance | 1.02 | | | | | | | | | | | | S80 |
| Thyroid | HypTens Asthma Dyspepsia COPD IrritBSynd | | 0.23 | | | | | | | | | | | S176 |
| Depres Alcohol | HypTens PainCond Diabetes Anxiety Substance | | | 0.51 | | | | | | | | | | S256 |
| HypTens CHD | PainCond Diabetes Arthritis StrokeTIA HeartFail | | | | 0.98 | 2.39 | 2.49 | 3.12 | | | | | | S207 |
| Asthma | Depres PainCond Dyspepsia COPD Anxiety | | | | | 1.10 | | | | | | | | S226 |
| Depres | PainCond CHD Anxiety IrritBSynd Substance | | | | | | 0.27 | | | | | | | S26 |
| PainCond Anxiety | HypTens Depres Dyspepsia COPD Consti | | | | | | 1.14 | 0.86 | | | | | | S85 |
| Depres CHD | HypTens PainCond Diabetes COPD Anxiety | | | | | | 1.30 | | | | | | | S290 |
| HypTens PainCond | Depres CHD Diabetes Arthritis CKD | | | | | | | 2.12 | 2.98 | 2.84 | 4.65 | | | S49 |
| CHD | HypTens PainCond Thyroid IrritBSynd Diverticlr | | | | | | | 1.27 | | | | | | S19 |
| Alcohol | HypTens CHD Diabetes Arthritis COPD | | | | | | | 0.50 | | | | | | S169 |
| HypTens Asthma CHD COPD | Depres PainCond Diabetes CKD HeartFail | | | | | | | 0.18 | | | | | | S1 |
| Dyspepsia | HypTens Depres IrritBSynd Consti Diverticlr | | | | | | | 1.50 | | | | | | S181 |
| HypTens Diabetes | Depres PainCond Thyroid Substance CKD | | | | | | | 1.15 | | | | | | S337 |
| HypTens | PainCond Asthma Dyspepsia Diabetes Arthritis | | | | | | | 3.95 | | | | | | S59 |
| PainCond COPD | HypTens Depres Dyspepsia Anxiety Consti | | | | | | | | 1.39 | | | | | S86 |
| Prostat | HypTens CHD Arthritis HearImpr Diverticlr | | | | | | | | 2.04 | | | | | S240 |
| HypTens Depres | Diabetes Thyroid StrokeTIA CKD SchizBipol | | | | | | | | 0.87 | | | | | S29 |
| CHD StrokeTIA AtrFib | HypTens PainCond Dyspepsia Substance HeartFail | | | | | | | | | 0.72 | | | | S267 |
| CHD Substance | HypTens Depres PainCond Consti Dementia | | | | | | | | | 1.07 | | | | S155 |
| HypTens Diverticlr | CHD Consti StrokeTIA CKD AtrFib | | | | | | | | | | 2.95 | | | S162 |
| HypTens AtrFib | CHD Diabetes StrokeTIA CKD HeartFail | | | | | | | | | | 2.59 | | | S193 |
| Consti | Depres PainCond Dyspepsia Anxiety Dementia | | | | | | | | | | 1.88 | | | S221 |
| IrritBSynd | HypTens PainCond HearImpr Consti Diverticlr | | | | | | | | | | | 0.97 | | S163 |

| | | | | | | | | | | | |
|---|---|---|---|---|---|---|---|---|---|---|---|
| HypTens Substance StrokeTIA | CHD Thyroid Arthritis Consti AtrFib | | | | | | | | | 0.53 | S269 |
| HypTens StrokeTIA | HearImpr COPD AtrFib PeriVascDis HeartFail | | | | | | | | | 0.42 | S105 |
| CHD Substance Consti | Depres PainCond Thyroid Anxiety Dementia | | | | | | | | | 1.59 | S32 |
| Asthma HearImpr COPD | Depres Dyspepsia Thyroid Anxiety Consti | | | | | | | | | 0.39 | S63 |
| Depres Dyspepsia | Thyroid Anxiety Consti Diverticlr Dementia | | | | | | | | | 0.55 | S173 |
| HypTens Substance Diverticlr | CHD Dyspepsia Cancer Consti StrokeTIA | | | | | | | | | 0.59 | S174 |
| Anxiety Consti | HypTens Depres HearImpr Glaucoma VisualImpr | | | | | | | | | 1.43 | S341 |
| Substance | HypTens CHD Consti CKD Dementia | | | | | | | | | 3.61 | S151 |
| CHD Anxiety | HypTens Depres HearImpr CKD Diverticlr | | | | | | | | | 1.74 | S143 |
| Diverticlr | Asthma Thyroid StrokeTIA HeartFail VisualImpr | | | | | | | | | 0.59 | S326 |
| StrokeTIA | HypTens CHD Diabetes HeartFail Dementia | | | | | | | | | 1.64 | S12 |

Supplement Table 7: First cluster (i.e., the representative) of every cluster group with A) seven (upper half of the table) and B) more than seven (lower half of the table) borderline condition(s) (i.e., prevalence is between 0.3 and 0.7) and more than 30 conditions of low prevalence (i.e., prevalence is below 0.3).

| LTC (prevalence > 0.7) | Moderate LTCs | Cluster size in age groups (% of age group) | | | | | | | | | | | | Cluster Set |
|---|---|---|---|---|---|---|---|---|---|---|---|---|---|---|
| | | 1 | 2 | 3 | 4 | 5 | 6 | 7 | 8 | 9 | 10 | 11 | 12 | |
| Depres PainCond | HypTens Asthma Diabetes Anxiety IrritBSynd Substance Consti | | 0.31 | | | | | | | | | | | S81 |
| PainCond | HypTens Depres Asthma CHD Arthritis COPD IrritBSynd | | | 0.67 | | | | | | | | | | S118 |
| Depres IrritBSynd | HypTens PainCond Asthma COPD Anxiety Consti Diverticlr | | | | | | 0.86 | | | | | | | S27 |
| Dyspepsia | HypTens Depres PainCond Anxiety IrritBSynd Consti Diverticlr | | | | | | | | 2.25 | | | | | S182 |
| Dementia | HypTens CHD HearImpr COPD Diverticlr AtrFib HeartFail | | | | | | | | | | 0.89 | | | S149 |
| Consti | Depres PainCond Dyspepsia COPD Anxiety IrritBSynd Diverticlr | | | | | | | | | | 1.80 | | | S184 |
| HypTens HearImpr | Depres Diabetes COPD Substance StrokeTIA AtrFib Dementia | | | | | | | | | | 0.14 | | | S242 |
| HypTens CHD | PainCond Diabetes Thyroid Consti StrokeTIA AtrFib HeartFail | | | | | | | | | | | 3.42 | | S114 |
| HypTens CKD | Depres PainCond Thyroid Arthritis Anxiety Consti HeartFail | | | | | | | | | | | 1.05 | | S203 |
| Depres Substance | Diabetes Thyroid HearImpr COPD Anxiety Diverticlr HeartFail | | | | | | | | | | | | 0.48 | S98 |
| LTC (prevalence > 0.7) | Moderate LTCs | 1 | 2 | 3 | 4 | 5 | 6 | 7 | 8 | 9 | 10 | 11 | 12 | |
| Depres | HypTens PainCond CHD Diabetes Anxiety Substance Consti StrokeTIA | | | | | | | | 0.57 | | | | | S219 |
| COPD | HypTens Depres PainCond Asthma CHD Anxiety Consti HeartFail | | | | | | | | | 1.00 | 1.95 | | | S277 |
| HeartFail | Depres CHD COPD Alcohol Substance StrokeTIA AtrFib Dementia | | | | | | | | | 0.48 | | | | S148 |
| Diverticlr | HypTens Depres PainCond CHD Thyroid Anxiety IrritBSynd Consti | | | | | | | | | 1.34 | | | | S323 |
| Dyspepsia | HypTens Depres PainCond Anxiety Cancer Substance Consti Diverticlr | | | | | | | | | 0.77 | | | | S183 |
| Substance | Depres CHD Thyroid COPD Consti StrokeTIA AtrFib Dementia | | | | | | | | | | | 0.59 | | S33 |

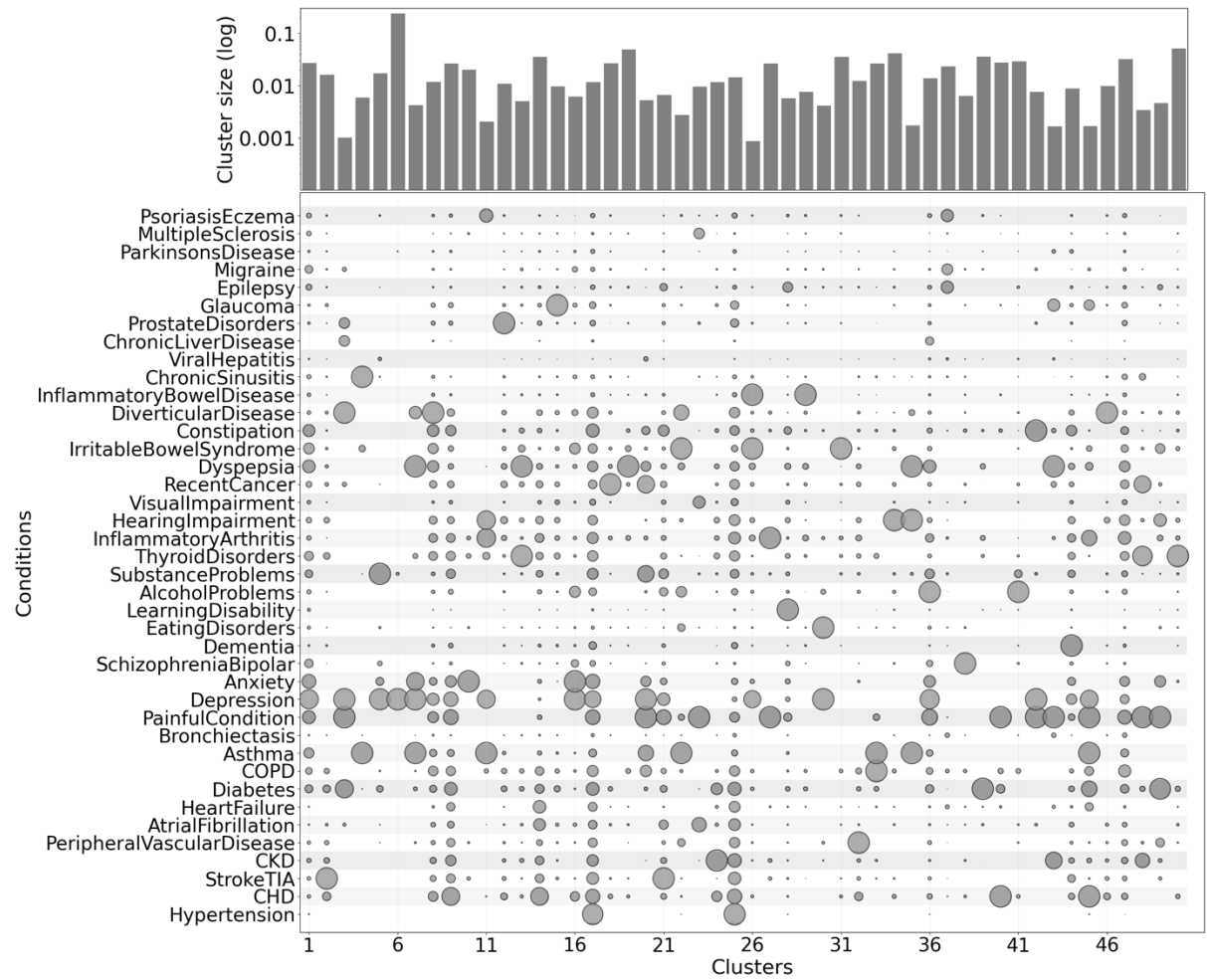

Supplement Figure 7: Cluster prevalence for best LCA run (K = 50) on the whole data (i.e. people aged 40 years and over who have at least one LTC).